\documentclass[journal]{IEEEtran}
\usepackage{amsmath,amsfonts}
\usepackage{algorithmic}
\usepackage{array}
\usepackage[caption=false,font=small,labelfont=sf,textfont=sf]{subfig}
\usepackage{stfloats}
\usepackage{pgfplots}
\usepackage{url}
\usepackage{verbatim}
\usepackage{graphicx}
\usepackage{enumitem}
\usepackage{fix-cm}
\usepackage{pgfplots}
\def\BibTeX{{\rm B\kern-.05em{\sc i\kern-.025em b}\kern-.08em
    T\kern-.1667em\lower.7ex\hbox{E}\kern-.125emX}}
\usepackage{balance}

\usepackage{cite}           % correct
\usepackage[colorlinks=true,linkcolor=blue,citecolor=blue,urlcolor=blue]{hyperref}

\usepackage{algorithmic}
\usepackage{textcomp}
\usepackage{xcolor}
\usepackage{multirow} 
\usepackage{amssymb,amsfonts,amsmath,graphicx,tikz,textcomp}
\usepackage{booktabs}
\usepackage{flushend}
\usepackage{amssymb}
\usetikzlibrary{shapes,arrows}
\usetikzlibrary{positioning} 
\usepackage{psfrag}
\usepackage{makecell}
\usepackage{diagbox}
\usepackage{siunitx}
\usepackage{bm}

\usepackage{acronym}
\acrodef{DNN}{deep neural network}
\acrodef{GAN}{generative adversarial network}
\acrodef{DisCoGAN}{discriminatively conditoned GAN}
\acrodef{SOTA}{state-of-the-art}
\acrodef{SE}{Speech enhancement}
\acrodef{SR}{speech restoration}
\acrodef{WER}{word error rate}
\acrodef{CER}{character error rate}
\acrodef{ASR}{automatic speech recognition}
\acrodef{E2E}{end-to-end}
\acrodef{TF}{time-frequency}
\acrodef{STFT}{short-time Fourier transform }
\acrodef{GAN}{generative adversarial network}
\acrodef{RIR}{room impulse response }
\acrodef{FiLM}{feature-wise linear modulation }
\acrodef{SNR}{signal-to-noise ratio}
\acrodef{MOS}{mean opinion score}
\acrodef{DNN}{deep neural network}
\acrodef{PDF}{probability density function}
\acrodef{Conv}{convolution}

\def\vecx{\mathbf{x}}
\def\vecs{\mathbf{s}}
\def\vecv{\mathbf{v}}
\def\R#1{\mathbb{R}^{#1}}
\def\hats{\hat{\mathbf{s}}}
\def\noiseprior{p(\mathbf{z})}
\def\noisesample{\mathbf{z}}
\def\gen{\mathcal{G}_{\theta}}
\def\px{p_x}
\def\ps{p_s}
\def\hatsdisc{\hat{\mathbf{s}}_{\text{disc}}}
\def\phatsdisc{p_{\hat{\mathbf{s}}_{\text{disc}}}}
\newcommand{\Dl}{\mathbf{D}_{\text{L}}}
\newcommand{\FTheta}{\mathcal{F}_{\Theta}}
\newcommand{\FThetEnc}{\mathcal{F}_{\Theta_{\text{enc}}}}
\newcommand{\ThetaEnc}{\Theta_{\text{enc}}}
\newcommand{\varphiParam}{\varphi}

\newcommand{\LD}{\mathcal{L}_\text{D}}
\newcommand{\LG}{\mathcal{L}_\text{G}}
\newcommand{\Lrec}{\mathcal{L}_{\text{rec}}}
\newcommand{\Ladv}{\mathcal{L}_{\text{adv}}}
\newcommand{\Lfeat}{\mathcal{L}_{\text{feat}}}
\newcommand{\Lt}{\mathcal{L}_\text{t}}
\newcommand{\Lf}{\mathcal{L}_\text{f}}

\def\lambdat{{\lambda_\text{t}}}
\def\lambdaf{{\lambda_\text{f}}}

\newcommand{\Stft}{\mathbf{S}}          % STFT matrix of clean speech
\newcommand{\hatStft}{\hat{\mathbf{S}}} % STFT matrix of estimate
\newcommand{\complexMat}[2]{\mathbb{C}^{#1 \times #2}}
\newcommand{\RealMat}[2]{\mathbb{R}^{#1 \times #2}}
\newcommand{\Qset}{\mathcal{Q}}

\newcommand{\Pow}{\mathbf{P}}   % Power spectrum
\newcommand{\Mel}{\mathbf{M}}   % Mel spectrum

\newcommand{\discr}{\mathcal{D}_\phi}
\newcommand{\hinge}[1]{\left[ #1 \right]_+}
\newcommand{\disc}{\mathcal{D}}
\newcommand{\DNN}{\mathcal{U}}
\newcommand{\Gl}{\mathbf{G}_\text{L}}
\newcommand{\Gdl}{\mathbf{G}_\text{DL}}
\newcommand{\Zl}{\mathbf{Z}_\text{L}}
\newcommand{\Stftx}{\mathbf{X}}  
\newcommand{\Stftw}{\mathbf{W}}  
\newcommand{\Stfty}{\mathbf{Y}}  
\newcommand{\Hout}{\mathbf{H}} 
\newcommand{\concat}[1]{\mathrm{concat}\left( #1 \right)}
\newcommand{\DecFeat}{\mathbf{D_\text{f}}} % Decoder features
\newcommand{\EncFeat}{\mathbf{E_\text{f}}} % Encoder features
\newcommand{\Attn}{\mathbf{A}}      % Attention weights
\newcommand{\Conv}{\mathrm{Conv}}   % Convolution
\newcommand{\ReLU}{\mathrm{ReLU}}   % ReLU
\newcommand{\LSTM}{\mathrm{LSTM}}
\newcommand{\Sigmoid}{\mathrm{Sigmoid}} % Sigmoid
\newcommand{\RealTensor}[3]{\mathbb{R}^{#1 \times #2 \times #3}} % 3D tensor shape
\newcommand{\dg}{d_\text{g}}
\newcommand{\dd}{d_\text{d}}
\newcommand{\blockdiag}{\mathrm{blockdiag}}
\newcommand{\tildedl}{\widetilde{\mathbf{D}}_{\text{L}}}

\begin{document}
\title{Leveraging Discriminative Latent Representations for \\ Conditioning GAN-Based Speech Enhancement}

\author{Shrishti Saha Shetu, 
        Emanu\"{e}l A. P. Habets, \IEEEmembership{Senior Member,~IEEE}, 
        Andreas Brendel, \IEEEmembership{Member,~IEEE}%
\thanks{The authors are with the International Audio Laboratories Erlangen, a joint institution of the Friedrich-Alexander-Universität Erlangen-Nürnberg (FAU) and Fraunhofer IIS, 91058 Erlangen, Germany (e-mail: shrishti.saha.shetu@iis.fraunhofer.de; emanuel.habets@audiolabs-erlangen.de; andreas.brendel@iis.fraunhofer.de).}}

%\markboth{Journal of \LaTeX\ Class Files,~Vol.~18, No.~9, September~2020}%
%{How to Use the IEEEtran \LaTeX \ Templates}

\maketitle

\begin{abstract}
Generative speech enhancement methods based on generative adversarial networks (GANs) and diffusion models have shown promising results in various speech enhancement tasks. However, their performance in very low signal-to-noise ratio (SNR) scenarios remains under-explored and limited, as these conditions pose significant challenges to both discriminative and generative state-of-the-art methods. To address this, we propose a method that leverages latent features extracted from discriminative speech enhancement models as generic conditioning features to improve GAN-based speech enhancement. The proposed method, referred to as DisCoGAN, demonstrates performance improvements over baseline models, particularly in low-SNR scenarios, while also maintaining competitive or superior performance in high-SNR conditions and on real-world recordings. We also conduct a comprehensive evaluation of conventional GAN-based architectures,  including GANs trained end-to-end, GANs as a first processing stage, and post-filtering GANs, as well as discriminative models under low-SNR conditions. We show that DisCoGAN consistently outperforms existing methods. Finally, we present an ablation study that investigates the contributions of individual components within DisCoGAN and analyzes the impact of the discriminative conditioning method on overall performance.
\end{abstract}


\section{Introduction}
\label{Intro}
\ac{SE} aims to improve the quality and intelligibility of speech by reducing noise and other distortions present in observed speech signals \cite{boll2003suppression,paliwal2012speech,cohen2003noise,cohen2001speech}. In recent years, significant progress has been made in the \ac{SE} domain through the use of various \ac{DNN}-based approaches, leading to significant improvements in both speech quality and intelligibility. Until recently, most \ac{DNN}-based \ac{SE} methods relied on discriminative training techniques \cite{hu2020dccrn,liu2023mask,schroter2022deepfilternet2,choi2021real,zhao2022frcrn,shetu2023ultra,wang2018supervised,luo2018tasnet}. These methods have shown impressive performance in moderate to high \ac{SNR} conditions and have been benchmarked across numerous standard datasets \cite{reddy2020interspeech,valentini2016investigating}. However, recent studies \cite{hao2020masking,shetu2024comparative} have shown that under very low \ac{SNR} conditions, where the desired speech is often heavily masked by dominant noise components, many of these \ac{SOTA} discriminative \ac{SE} methods are unable to effectively suppress noise without also distorting or suppressing the speech, leading to a significant decline in overall speech quality.

In contrast, generative \ac{SE} approaches promise the potential for superior performance in these scenarios by learning the distribution of clean speech signals. This allows them to generate the desired speech signal by conditioning it on its noisy counterpart. However, most \ac{SOTA} generative \ac{SE} methods are designed for moderate \ac{SNR} conditions and can be grouped into two main categories: \ac{GAN}-based methods \cite{pascual2017segan,fu2021metricgan+,cao2022cmgan,strauss2023sefgan,chen2025tfdense,elgiriyewithana2024comprehensive} and, more recently, diffusion-based techniques \cite{richter2023speech, lu2022conditional, lemercier2023storm,yen2023cold,zhao2025conditional}. Recent diffusion-based approaches have shown effectiveness in improving speech quality in various adverse acoustical conditions, including low \ac{SNR} scenarios \cite{scheibler24_interspeech, wang2023cross}. However, diffusion models typically require batch processing with multiple reverse diffusion steps during inference to achieve high-quality \ac{SE}, which limits their applicability in frame-by-frame, causal, or semi-causal real-time processing scenarios. 

To this end, \ac{GAN}-based methods dominate practical applications, as they do not impose significant constraints on the \ac{SE} model in training or inference: The generator in a \ac{GAN} model can be trained \ac{E2E} with an adversarial loss and can be deployed for inference similarly to discriminative methods, enabling efficient real-time inference \cite{cao2022cmgan,chen2025tfdense}. As a result, \ac{GAN}s have been also widely applied in many speech reconstruction tasks \cite{ristea2024icassp}, including noise reduction, packet loss concealment, declipping, and speech inpainting. These problems are closely related to \ac{SE} in very low SNR conditions, where substantial portions of the speech signal may be lost due to additive noise and other degradations, requiring generative modeling to reconstruct the clean speech signal. Most \ac{SOTA} systems for speech reconstruction employ a two-stage architecture that integrates both generative and discriminative models. In the typical configuration, referred to here as \ac{GAN}-first, a \ac{GAN}-based restoration module is followed by a discriminative enhancement or post-filtering module \cite{10625840, yu2024ks, 10626452}. Alternatively, in post-filtering \ac{GAN}s, referred to here as \ac{GAN}-last, a \ac{GAN} is applied after a discriminative or traditional signal processing method \cite{kaneko2017generative, korse2022postgan,serbest2025deepfiltergan}. However, none of these two-stage \ac{GAN}-first and \ac{GAN}-last-based methods facilitate the flow of intermediate or latent representations between processing modules. These methods either operate directly on the noisy input or solely on the output of prior stages, without leveraging latent feature representations.

In our previous work, we addressed this limitation by deriving conditioning information for generative \ac{SE} from a surrogate discriminative \ac{SE} task \cite{shetu2025gan}. The effectiveness of this approach was demonstrated through a \ac{GAN}-based \ac{SE} system designed for low \ac{SNR} scenarios. The motivation for this approach is that the discriminative model’s ability to separate speech from noise should make the features extracted in its bottleneck layer highly informative for conditioning the \ac{GAN}, thereby improving its \ac{SE} performance. Compared to conventional \ac{GAN}-first or \ac{GAN}-last methods, our approach also simplifies the overall system by allowing the discriminative and generative models to run in parallel, eliminating the need to use the decoder of the discriminative model during training of the \ac{GAN} model and inference. In this work,  our specific contributions are as follows:
\begin{itemize}[labelindent=2pt, leftmargin=!]
    \item We propose the use of latent feature representations from discriminative models as generic conditioning features for GAN-based \ac{SE}, referred to as \ac{DisCoGAN}. The effectiveness of this approach is validated using features extracted from various pretrained discriminative \ac{SE} models. Furthermore, we propose a time-frequency-domain SEANet-based generator architecture \cite{tagliasacchi2020seanetmultimodalspeechenhancement, du2023funcodecfundamentalreproducibleintegrable} tailored for the DisCoGAN framework.
    
    \item We conduct a comprehensive evaluation of existing GAN-based approaches—namely \ac{E2E}, \ac{GAN}-first, and \ac{GAN}-last methods under low \ac{SNR} conditions, and compare their performance with our proposed DisCoGAN method.
    
    \item We assess the effectiveness of the DisCoGAN method in both low and high \ac{SNR} conditions by benchmarking its performance on various standard synthetic and real-world recording datasets, and compare it with many discriminative and generative baselines.
    
    \item We perform an ablation study to investigate the contribution of different components within the DisCoGAN framework, including the impact of the discriminative latent feature conditioning method.
\end{itemize}

In this work, we show that in low \ac{SNR} scenarios, the proposed \ac{DisCoGAN} outperforms conventional \ac{GAN}-based methods as well as baseline generative and discriminative models across all intrusive and non-intrusive objective and subjective metrics. In high \ac{SNR} scenarios, DisCoGAN achieves comparable or superior results relative to baseline \ac{GAN}-based and diffusion-based methods. We further demonstrate that \ac{DisCoGAN} generalizes well to real environments by benchmarking its performance on real test recordings, where it also achieves comparable performance to baseline methods.

The remainder of this paper is structured as follows. Sections~\ref{PF} and~\ref{PM} present the problem formulation and the proposed methods. Section~\ref{DA} details the overall DisCoGAN architecture, covering the generator model, discriminative conditioning approach, discriminator architecture, and a brief overview of the various discriminative models used to extract latent conditioning features. Section~\ref{ED} describes the training and evaluation datasets, objective evaluation metrics, different DisCoGAN variants, and baseline methods. Section~\ref{results} presents the objective and subjective results obtained in different experimental scenarios, followed by the ablation study. Finally, the conclusion is provided in Sec.~\ref{conclusions}.

\section{Problem Formulation}
\label{PF}
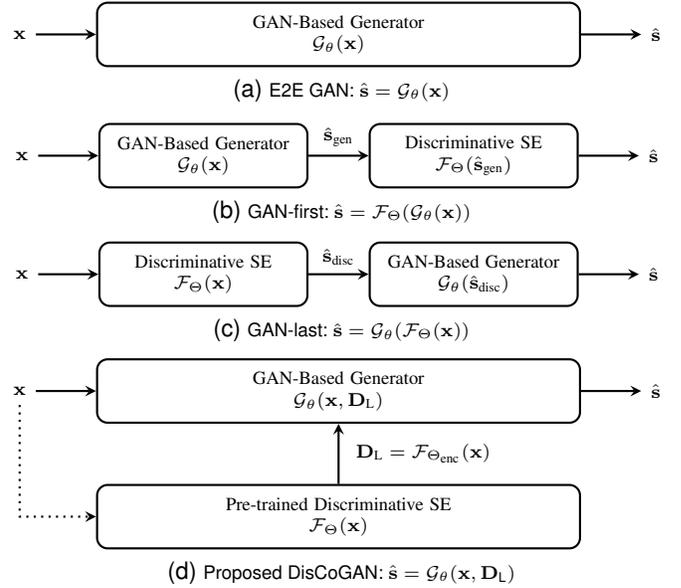
\begin{figure}
    \centering

    \subfloat[\scriptsize \ac{E2E} GAN: $\hats = \gen(\vecx)$]{%
        \begin{minipage}[b]{\linewidth}
            \centering
            \begin{tikzpicture}[>=stealth, thick, node distance=0.8cm, font=\scriptsize]
                \node (x4) {$\vecx$};
                \node[rectangle, draw=black, right=of x4, text width=17.6em, text centered, rounded corners, minimum height=2.4em] (gen4) {GAN-Based Generator \\ $\gen(\vecx)$};
                \node[right=of gen4] (s4) {$\hats$};

                \draw[->] (x4) -- (gen4);
                \draw[->] (gen4) -- (s4);
            \end{tikzpicture}
        \end{minipage}
        \label{subfig:E2E}
    }
    \vspace{-.4em}
    \subfloat[\scriptsize GAN-first: $\hats = \FTheta(\gen(\vecx))$]{%
        \begin{minipage}[b]{\linewidth}
            \centering
            \begin{tikzpicture}[>=stealth, thick, node distance=0.8cm, font=\scriptsize]
                \node (x1) {$\vecx$};
                \node[rectangle, draw=black, right=of x1, text width=7.2em, text centered, rounded corners, minimum height=2.4em] (gen1) {GAN-Based Generator \\ $\gen(\vecx)$};
                \node[rectangle, draw=black, right=of gen1, text width=7.2em, text centered, rounded corners, minimum height=2.4em] (dis1) {Discriminative SE \\ $\FTheta(\hats_{\text{gen}})$};
                \node[right=of dis1] (s1) {$\hats$};

                \draw[->] (x1) -- (gen1);
                \draw[->] (gen1) -- node[above] {\scriptsize $\hats_{\text{gen}}$} (dis1);
                \draw[->] (dis1) -- (s1);
            \end{tikzpicture}
        \end{minipage}
        \label{subfig:two-stage}
    }
    \vspace{-.4em}
    \subfloat[\scriptsize GAN-last: $\hats = \gen(\FTheta(\vecx))$]{%
        \begin{minipage}[b]{\linewidth}
            \centering
            \begin{tikzpicture}[>=stealth, thick, node distance=0.8cm, font=\scriptsize]
                \node (x2) {$\vecx$};
                \node[rectangle, draw=black, right=of x2, text width=7.2em, text centered, rounded corners, minimum height=2.4em] (dis2) {Discriminative SE \\ $\FTheta(\vecx)$};
                \node[rectangle, draw=black, right=of dis2, text width=7.2em, text centered, rounded corners, minimum height=2.4em] (gen2) {GAN-Based Generator \\ $\gen(\hats_{\text{disc}})$};
                \node[right=of gen2] (s2) {$\hats$};

                \draw[->] (x2) -- (dis2);
                \draw[->] (dis2) -- node[above] {\scriptsize $\hats_{\text{disc}}$} (gen2);
                \draw[->] (gen2) -- (s2);
            \end{tikzpicture}
        \end{minipage}
        \label{subfig:postfiltering}
    }
    \vspace{-.4em}
    \subfloat[\scriptsize Proposed DisCoGAN: $\hats = \gen(\vecx, \Dl)$]{%
        \begin{minipage}[b]{\linewidth}
            \centering
            \begin{tikzpicture}[>=stealth, thick, node distance=0.8cm and 0.8cm, font=\scriptsize]
                \node (x3) {$\vecx$};
                \node[rectangle, draw=black, right=of x3, text width=17.6em, text centered, rounded corners, minimum height=2.4em] (gen3) {GAN-Based Generator \\ $\gen(\vecx, \Dl)$};
                \node[rectangle, draw=black, below=of gen3, text width=17.6em, text centered, rounded corners, minimum height=2.4em] (dis3) {Pre-trained Discriminative SE \\ $\FTheta(\vecx)$};

                \node[anchor=west, below right=1cm and .1cm of gen3.north] (dl_equation) {$\Dl = \FThetEnc(\vecx)$};
                \node[right=of gen3] (s3) {$\hats$};

                \draw[->] (x3) -- (gen3);
                \draw[dotted, thick, ->] (x3) |- (dis3);
                \draw[->] (dis3) -- (gen3);
                \draw[->] (gen3) -- (s3);
            \end{tikzpicture}
        \end{minipage}
        \label{subfig:discogan}
    }
    \caption{Illustration of existing \ac{GAN}-based approaches and the proposed DisCoGAN method.}
    \label{Fig:GAN-based-configurations}
%\vspace{-1em}
\end{figure}

We assume an additive signal model, where the noisy signal \( \vecx \in \R{N} \) is the sum of the clean speech \( \vecs \in \R{N} \) and noise \( \vecv \in \R{N} \), where \( N \) denotes the length of the signal in samples. In classical conditional \ac{GAN}-based \ac{SE} methods, the objective is to generate an estimate \( \hats \) of the clean speech signal \( \vecs \), given the conditional information and the noise prior \( \noiseprior \). However, recent studies have shown that during \ac{GAN} training, the noise prior \( \noiseprior \) is effectively ignored when strong conditional information is provided \cite{isola2017image}. Hence, realizations \( \noisesample \) of the noise prior \( \noiseprior \) are not utilized as additional input in the proposed \cite{fu2019metricgan, fu2021metricgan+} \ac{GAN} framework. 

In the literature, \ac{E2E} \ac{GAN}-based \ac{SE} methods \cite{cao2022cmgan, fu2019metricgan, fu2021metricgan+}, as well as two-stage \ac{GAN}-first methods for speech reconstruction tasks \cite{tagliasacchi2020seanetmultimodalspeechenhancement, du2023funcodecfundamentalreproducibleintegrable}, typically use the noisy speech signal \( \vecx\) as the only conditioning information. In this approach, the generator \( \gen \) (where $\theta$ represents the parameters of the generator model), as  illustrated in Figs.~\ref{subfig:E2E} and \ref{subfig:two-stage}, implicitly learns the marginal \ac{PDF} of the clean speech signal \(\vecs\), given by
\begin{equation}
p_G(\vecs) = \int \px(\vecx) \, \delta(\vecs - \gen(\vecx)) \, d\vecx,
%\vspace{-.1cm}
\end{equation}
where \(\delta(\cdot)\) denotes the Dirac delta distribution, and \( p_x(\vecx) \) is the \ac{PDF} of the noisy signal \(\vecx\).

In contrast, \ac{GAN}-last methods, as illustrated in Fig.~\ref{subfig:postfiltering}, utilize an estimated clean speech signal \( \hatsdisc \), obtained from a discriminative \ac{SE} module or other speech synthesis approaches, as conditioning information \cite{korse2022postgan, kaneko2017generative, 7953090, sani2023improving}. These methods learn the marginal \ac{PDF} in this form
\begin{equation}
p_G(\vecs) = \int \phatsdisc(\hatsdisc) \, \delta\big(\vecs- \gen(\hatsdisc)\big) \, d\hatsdisc,
%\vspace{-.1cm}
\end{equation}
where \( \phatsdisc(\hatsdisc) \) denotes the \ac{PDF} of the clean speech estimated by the discriminative model. 

Recent studies \cite{10626452, serbest2025deepfiltergan}, have shown that the two-stage \ac{GAN}-first and \ac{GAN}-last (Fig.~\ref{subfig:two-stage} and Fig.~\ref{subfig:postfiltering}) methods often outperform \ac{E2E} approaches (Fig.~\ref{subfig:E2E}). However, existing two-stage approaches do not facilitate the exchange of intermediate or latent information between generative and discriminative modules. These design choices can render the second stage of the processing chain under-utilized in certain scenarios and entirely dependent on the conditioning signals $\hats_{\text{gen}}$ and $\hats_{\text{disc}}$, obtained from the first processing stage of GAN-first and GAN-last methods, respectively. We hypothesize that, in very low \ac{SNR} scenarios, where the first stage fails to generate a reliable clean speech estimates $\hats_{\text{gen}}$ and $\hats_{\text{disc}}$ from its noisy prior \( p(\vecx) \), the second stage cannot contribute effectively. In such scenarios, the existing two-stage methods  may result in suboptimal performance in very low \ac{SNR} scenarios \cite{shetu2025gan}.

\section{Proposed Method}
\label{PM}

In this work, we leverage the latent features extracted from the bottleneck of a discriminative \ac{SE} model as conditioning features for \ac{GAN}-based \ac{SE}. The discriminative model's ability to distinguish between speech and noise yields encoded bottleneck features that may differ from those learned by the \ac{GAN} encoder itself. Since these features are derived from a dedicated discriminative objective, it renders particularly valuable for conditioning the \ac{GAN}, thereby enhancing the performance of the generative \ac{SE} method \cite{shetu2025gan}. The conditioning information for the \ac{GAN}-based \ac{SE} model is learned through a surrogate discriminative \ac{SE} model, as illustrated in Fig.~\ref{subfig:discogan}. This design also enables a parallel model architecture where the discriminative model’s decoder can be discarded during \ac{GAN} training and inference, reducing overall computational complexity. We refer to the proposed method that conditions a \ac{GAN} with such discriminatively learned features as \ac{DisCoGAN}. 

Let $\Dl$ denote the output of the encoder of a pre-trained discriminative \ac{SE} model \(\FTheta\), such that $\Dl = \FThetEnc(\vecx)$, where \(\ThetaEnc\)   represents the encoder parameters of the discriminative model \(\mathcal{F}\). The generator \(\gen\)  estimates the clean speech signal \(\hat{\vecs} = \gen(\vecx, \Dl)\) and learns to approximate the marginal \ac{PDF} of the clean speech signal $\vecs$,
\begin{equation}
    p_G(\vecs) = \iint p(\vecx, \Dl) \, \delta(\vecs - \gen(\vecx, \Dl)) \, d\vecx \, d\Dl,
%\vspace{-.1cm}
\end{equation}
where \( p(\vecx, \Dl) \) is the joint \ac{PDF} of the noisy input signal $\vecx$ and the latent features $\Dl$. The learning objective of the proposed conditional GAN can be expressed as
\begin{equation}
\begin{split}
\min_{\theta} \max_{\varphiParam} \; & \mathbb{E}_{\vecs \sim \ps} \left[ \LD(\varphiParam, \vecs) \right] + \mathbb{E}_{\vecx \sim \px} \left[ \LG(\varphiParam, \theta, \vecx, \Dl) \right],
\end{split}
%\vspace{-.1cm}
\end{equation}
where $\LD$ and $\LG$ denote the discriminator and generator loss functions, respectively, and $\varphiParam$ denotes the discriminator parameters.

%%%%%%%%%%%%%%%%%%%%%%%%%%%%%%%%%%%%%%%%%%%%%%%%

%%%Here, a different proposal for noting this down. In the notation above, you use Goodfellow's version of the GAN loss which is not the one you use later, which might be confusing. In this proposal, you can avoid specifying a GAN loss at this point by introducing to loss terms $\mathcal{L}$ that you can specify later
%%%%%%%%%%%%%%%%%%%%%%%%%%%%%%%%%%%%%%%%%%%%%%%%
\subsection{Learning Objectives}
\label{LO}
The training loss of 
\ac{DisCoGAN} consists of three components: reconstruction loss $\Lrec$, adversarial loss $\Ladv$, and feature matching loss $\Lfeat$. The reconstruction loss consists of time- and frequency-domain loss components $\Lt$ and $\Lf$, respectively. In the time domain a \(\ell_1\) loss is employed
\begin{equation}
    \Lt(\vecs, \hats) = \mathbb{E}_{\vecs, \hats} \|\vecs - \hats\|_1 .
%\vspace{-.1cm}
\end{equation}

In the frequency domain, both \(\ell_1\) and Frobenius distances are minimized on Mel and magnitude spectra at multiple resolutions. With \(\Stft_i, \hat{\Stft}_i \in \complexMat{F_i}{T_i}\) denoting the \ac{STFT} of the target and of the estimated signal $\vecs$ and $\hats$, respectively. The frequency-domain loss \(\Lf\) is given by
\begin{equation}
\begin{aligned}
\Lf(\vecs, \hats) = \mathbb{E}_{\vecs, \hats} \Bigg[ \frac{1}{|\Qset|} \sum_{i \in \Qset} \big(&
\|\Pow_i - \widehat{\Pow}_i\|_1 + \|\Pow_i - \widehat{\Pow}_i\|_\text{F} \\
&\hspace{-3em} + \|\Mel_i - \widehat{\Mel}_i\|_1 + \|\Mel_i - \widehat{\Mel}_i\|_\text{F} \big) \Bigg],
\end{aligned}
%\vspace{-.1cm}
\end{equation}
where $\Pow_i$  and $\Mel_i$ respectively denote the log power and Mel spectrograms of the target signal, and \(\|\cdot\|_\text{F}\) is the Frobenius norm. Multiple \ac{STFT} resolutions \(i \in \Qset = \{5, \ldots, 10\}\}\) use window sizes \(2^i\) and hop lengths of \(2^i/4\), with  \(|\Qset|\) denotes the number of resolutions. The overall reconstruction loss is $\Lrec  = \lambdat \Lt + \lambdaf \Lf$, where $\lambdat$ and $\lambdaf$ denote the weighting factors for the time-domain and frequency-domain loss components, respectively.

The adversarial loss contains the loss term \(\Ladv\) to train the generator \(\gen\), and the loss term $\LD$  to train the discriminator $\discr$:
\begin{equation}
\Ladv(\hat{\vecs}) = \mathbb{E}_{\vecx, \Dl} \left[ \frac{1}{K} \sum_{k,t=1}^{K,T_k} \frac{1}{T_k} \hinge{1 - \disc_{k,t}(\gen(\vecx, \Dl))} \right] ,
%\vspace{-.1cm}
\end{equation}
\begin{equation}
\begin{aligned}
\LD(\vecs, \hat{\vecs}) ={} & \mathbb{E}_{\vecs} \left[ \frac{1}{K T_k} \sum_{k,t=1}^{K,T_k} \hinge{1 - \disc_{k,t}(\vecs)} \right] \\
& \hspace{-2em} + \mathbb{E}_{\vecx, \Dl} \left[ \frac{1}{K T_k} \sum_{k,t=1}^{K,T_k} \hinge{1 + \disc_{k,t}(\gen(\vecx, \Dl))} \right] ,
\end{aligned}
%\vspace{-.1cm}
\end{equation}
where \(\left[ \vecx \right]_+ = \max(0, \vecx )\) denotes the hinge function,  $K$ is the number of discriminators  and \(\disc_{k,t}\) denotes the $k$th discriminator evaluated at time frame \(t\). The feature matching loss $\Lfeat$ is employed to measure the difference in intermediate features of the discriminator between target clean speech $\mathbf{s}$ and generated clean speech $\hat{\mathbf{s}}$
\begin{equation} 
\mathcal{L}_{\text{feat}}(\vecs, \hats) = \mathbb{E}_{\vecs, \hats} \left[ \frac{1}{KL} \sum_{k,t,l=1}^{K,T_k,L} \frac{1}{T_k} \left\| \disc^{(l)}_{k,t}(\vecs) - \disc^{(l)}_{k,t}(\hats) \right\|_1 \right] ,
%\vspace{-.1cm}
\end{equation}
Here, $L$ is the number of layers, and \(\disc^{(l)}_{k,t}\) represents the outputs of layer \(l\) of the $k$th discriminator evaluated at time step $t$. The total generator loss \(\LG\) is chosen as
\begin{equation} 
\mathcal{L}_G = \mathcal{L}_{rec} + \lambda_{\text{adv}} \mathcal{L}_{\text{adv}} + \lambda_{\text{feat}} \mathcal{L_{\text{feat}}} ,
%\vspace{-.1cm}
\end{equation}
where \(\lambda_{\text{adv}}\) and \(\lambda_{\text{feat}}\) denote the weighting factors for the adversarial loss and the feature matching loss, respectively.

\begin{figure*}[!t]
\centering
\includegraphics[width=.84\textwidth]{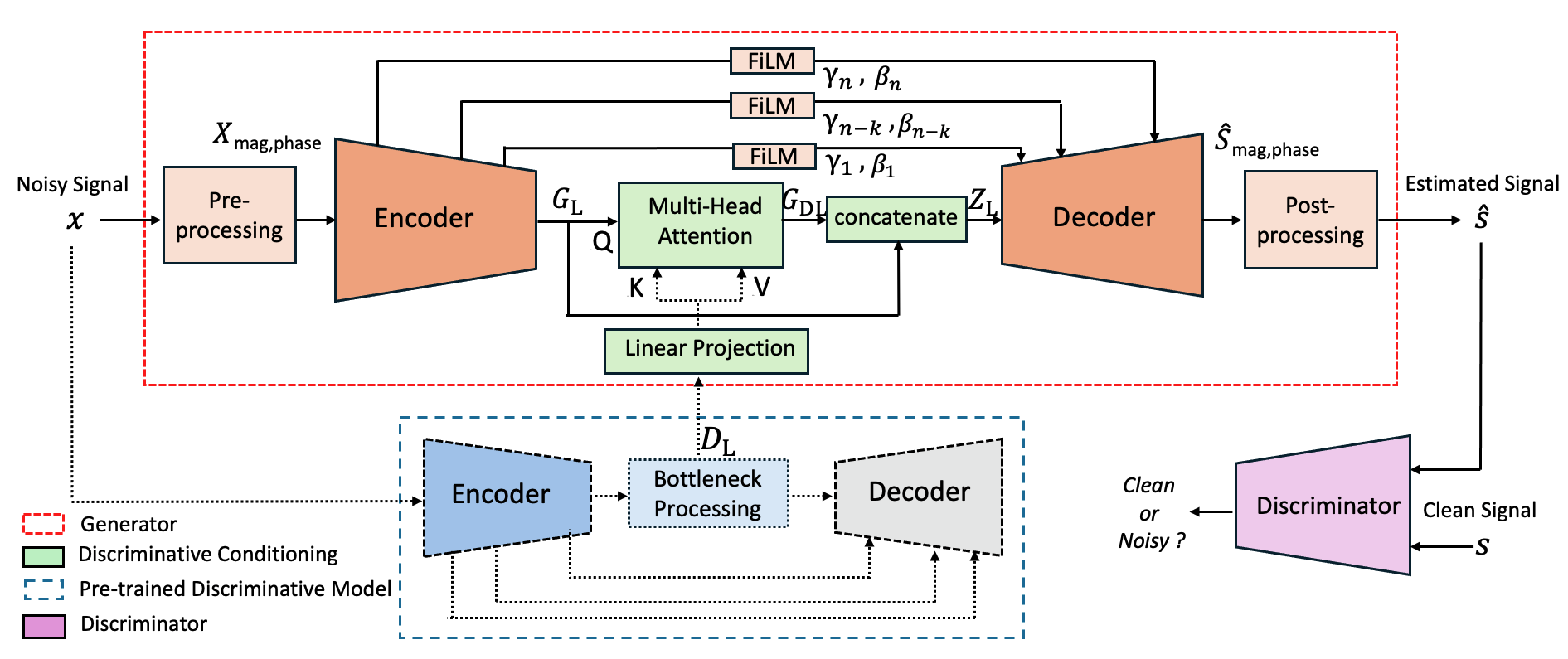}
\vspace{-.3em}
\caption{ Overview of  DisCoGAN: a \ac{TF}-domain SEANet-based \ac{GAN} conditioned with discriminative latent features. }
\label{fig:LF}
%\vspace{-1em}
\end{figure*}

\section{DisCoGAN Architecture}
\label{DA}
The DisCoGAN model (Fig.~\ref{fig:LF}) consists of three main components: i)~A generator learning the marginal \ac{PDF} of clean speech \(p(\vecs)\), ii)~a pre-trained discriminative \ac{SE} model extracting the conditioning features $\Dl$, and iii)~a discriminator for adversarial training of the generator. The discriminative and generative encoders process the noisy time-domain signal~$\vecx$. The output $\Dl$ of the discriminative encoder is used to adjust the output $\Gl$ of the generative encoder using a masked multi-head attention mechanism. The manipulated latent representation $\Gdl$ is stacked with $\Gl$, resulting in $\Zl$, which is then processed with the generative decoder to generate the estimate~$\hats$ of the clean speech signal~$\vecs$.

%\vspace{-1em}
\subsection{Generator Architecture} 
We use a SEANet-based model as the generator \cite{tagliasacchi2020seanetmultimodalspeechenhancement}, which has a UNet-like structure with a symmetric
encoder-decoder network with skip-connections. As the original SEANet operates in the time-domain, we adapt the architecture to the \ac{TF} domain by employing 2D \ac{Conv} layers in the encoder and decoder, following the approach in \cite{du2023funcodecfundamentalreproducibleintegrable}. In the pre-processing block, the noisy signal $\vecx$ is \ac{STFT}-transformed $\Stftx = \mathrm{STFT}(\vecx) \in \complexMat{F}{T} $,  and its log magnitude and phase $\Stftx_{{\text{mag},\text{phase}}} \in \RealTensor{3}{F}{T}$ are computed and concatenated in channel dimension, i.e.,
\begin{equation}
\begin{aligned}
\Stftx_{\textrm{mag,phase}} &= \concat{ \log(|\Stftx|), \frac{\Re(\Stftx)}{|\Stftx|}, \frac{\Im(\Stftx)}{|\Stftx|}} .
\end{aligned}
%\vspace{-.1cm}
\end{equation}
The generator $\gen$ outputs estimated clean speech spectral features $\hatStft_{{\text{mag},\text{phase}}}  \in \RealTensor{3}{F}{T}$, which consist of the magnitude and phase components $\left[ \hatStft_{\textrm{mag}}, \hatStft_{\textrm{r}}, \hatStft_{\textrm{i}} \right]$. The estimated time-domain speech signal $\hats$ is obtained as
\begin{equation}
\hat{\vecs} = \mathrm{ISTFT} \left( \mathrm{Softplus}(\hatStft_{\textrm{mag}}) \odot \left( \hatStft_{\textrm{r}} + j \hatStft_{\textrm{i}} \right) \right),
\end{equation}
where the inverse $\mathrm{STFT}$ is denoted as $\mathrm{ISTFT}$, $j=\sqrt{-1}$ is the imaginary unit, and $\odot$ is the Hadamard product. The $\mathrm{Softplus}(\cdot)$ activation is defined as $\mathrm{Softplus}(\vecx) = \log(1 + e^\vecx),$
which ensures non-negativity while retaining smooth gradients \cite{goodfellow2016deep}. Here,  the exponential is always applied element-wise.

\subsubsection{Encoder-Decoder} The encoder consists of a $2$D \ac{Conv} with $32$ channels with a kernel-size of $(2,4)$ followed by $8$ \ac{Conv} blocks,  each consisting of a single \ac{Conv} residual unit and a downsampling layer~\cite{du2023funcodecfundamentalreproducibleintegrable}. Downsampling is performed only in the frequency dimension via strided \ac{Conv} with a stride factor of $2$ per layer. Each residual unit comprises two \ac{Conv} layers with a kernel of size $(3 \times 3)$, a dilation factor of $2$ in the frequency dimension, and a skip connection. The channel number in the downsampling layers doubles with each block, up to a maximum of $C=512$ channels, reducing the frequency dimension to $F=1$ eventually, which results in an output shape of $(B, C=512,F=1,T)$. Here, $B$ and $T$ represent the batch size and number of frames, respectively. For simplicity of representation, the batch dimension is omitted in the following. These resulting features are processed with a two-layer LSTM of $512$ units for temporal sequence modeling and a final $1$D \ac{Conv} layer with $128$ output channels, representing the dimension of the generative models latent representation $\Gl \in \RealMat{128}{T}$. The decoder mirrors the encoder architecture, incorporating \ac{FiLM}-based skip connections between corresponding encoder and decoder \ac{Conv} blocks. Each decoder \ac{Conv} block uses transpose \ac{Conv} to upsample the frequency dimension by a factor of $2$ per layer, followed by a residual unit similar to the encoder architecture.

\subsubsection{FiLM Conditioning for Encoder-Decoder}
We implement a modified residual \ac{FiLM} layer~\cite{perez2018film} to condition the decoder outputs using the encoder features of the generator through skip connections.  In this way, we modulate the decoder features \(\DecFeat_n \in \RealTensor{C}{F}{T}\) at each \ac{Conv} block $n$ using the corresponding encoder features \(\EncFeat_n \in \RealTensor{C}{F}{T}\).  Each FiLM module computes a scale factor $\gamma_n = \ReLU\left(\Conv_\gamma(\EncFeat_n)\right)$ and a shift factor $\beta_n = \Sigmoid\left(\Conv_\beta(\EncFeat_n)\right)$ from the encoder features $\EncFeat_n$ using two separate $2$D \ac{Conv} with a kernel of size $(1 \times 3)$.
These factors are then multiplied by \ac{Conv} attention weights $\Attn_n \in \RealTensor{1}{F}{T} $ obtained from two sequential 2D \ac{Conv} with $\ReLU$ and $\Sigmoid$ activations, which first reduce the number of conditioning channels by a factor of \( 8 \) and then upsample it to the desired dimension with a pointwise convolution. The final FiLM-modulated decoder features are obtained using a residual affine transformation:
\begin{equation}
\begin{aligned}
\gamma_n' &= \gamma_n \odot \Attn_n, \quad \beta_n' = \beta_n \odot \Attn_n, \\
\DecFeat_n^{\mathrm{out}} &= \DecFeat_n + \left( \gamma_n' \odot \DecFeat_n + \beta_n' \right).
\end{aligned}
%\vspace{-.1cm}
\end{equation}

\subsubsection{Discriminative Conditioning}
\label{DiscCond}
We obtain the generative and discriminative latent features \(\Gl \in \RealMat{T}{\dg}\) and \(\Dl \in \RealMat{T}{\dd}\), respectively, from the encoder of the generator and a pretrained discriminative  \ac{SE} model. Here, \(\dg\) and \(\dd\) denote the latent feature dimensions. To condition the generative features \(\Gl\), we first align the time dimension of \(\Dl\) to match that of \(\Gl\) using linear interpolation \cite{chapra2011numerical}. Next, to match the feature dimension, we apply a linear transformation:
\begin{equation}
    \tildedl = \Dl \overline{\mathbf{W}}_\text{d} + \mathbf{B}_\text{d} \in \mathbb{R}^{T \times \dg},
\end{equation}
where \(\overline{\mathbf{W}}_\text{d} = \blockdiag(\mathbf{W}_\text{d}, .., .., \mathbf{W}_\text{d}) \in \RealMat{T \dd}{T \dg}\) and \(\mathbf{B}_\text{d}  =\blockdiag(\mathbf{b}_\text{d}, .., .., \mathbf{b}_\text{d}) \in \R{T \dg}\) are learnable block diagonal matrices. In the rest of the paper, for simplicity, we assume the time dimension \(T = 1\), reducing representation to standard weight matrix  \(\mathbf{W}_\text{d}\) and bias vector  \(\mathbf{b}_\text{d}\) forms. We then apply masked multi-head attention~\cite{nicolson2020masked} with a fixed lookahead of \(L = 20\) frames. The generator features \(\Gl \in \mathbb{R}^{T \times \dg}\) serve as queries, while the transformed discriminative features \(\widetilde{\mathbf{D}}_{\text{L}} \in \mathbb{R}^{T \times \dg}\) serve as keys and values:
\begin{equation}
    \mathbf{Q} = \Gl \mathbf{W}^\text{Q}, \quad
    \mathbf{K} = \widetilde{\mathbf{D}}_{\text{L}} \mathbf{W}^\text{K}, \quad
    \mathbf{V} = \widetilde{\mathbf{D}}_{\text{L}} \mathbf{W}^\text{V},
%\vspace{-.1cm}
\end{equation}
where \(\mathbf{W}^\text{Q}, \mathbf{W}^\text{K}, \mathbf{W}^\text{V} \in \mathbb{R}^{\dg \times \dg }\) are learnable projection matrices. The attention output is computed as
\begin{equation}
    \Gdl = \mathrm{Softmax}\left( \frac{\mathbf{Q} \mathbf{K}^\top}{\sqrt{\dg }} + \mathcal{M} \right) \mathbf{V},
\end{equation}
where the attention mask \(\mathcal{M} \in \{0, -\infty\}^{T \times T}\) is defined by
\begin{equation}
    \mathcal{M}_{p,q} = 
    \begin{cases}
    0, & q \in \{p, \min(p+L, T) \} \\
    -\infty, & \text{otherwise}
    \end{cases}.
%\vspace{-.1cm}
\end{equation}
We implement the masked multi-head attention mechanism with two attention heads using the PyTorch 2.3 default implementation. Finally, the attention output \(\Gdl\) is concatenated with the original generative features along the feature dimension to produce the conditioned representation:
\begin{equation}
    \Zl = \concat{\Gl, \Gdl}
 \in \mathbb{R}^{ T \times 2 \dg }.
%\vspace{-.1cm}
\end{equation}

%This final representation \(\mathbf{z_l}\) integrates contextual information from the discriminative model into the generative model, enhancing its representational capacity.

\subsection{Discriminative Models}
\label{Discriminative Models}
We experiment with $4$ groups of discriminative SE methods ( Tab.~\ref{tab:discriminativemodels}) to extract the discriminative features $\Dl$ based on 1)  estimating a complex-valued ideal ratio mask for attenuating noise components,  2) mapping the time-domain waveform,  3) decoupling the estimation of magnitude and phase components in multiple stages, and 4) mapping the complex-valued spectrum. As a representative for each of these groups, we investigate  the DCCRN~\cite{hu2020dccrn}, DDAEC~\cite{pandey2020densely}, TaylorSENet~\cite{ijcai2022p582} and GCRN~\cite{tan2019learning} methods. 

\begin{table}[!t]
\caption{\footnotesize Specifications of discriminative SE models.}
\centering
\small
%\vspace{-.2em}
\setlength{\tabcolsep}{4pt} % Adjusted padding slightly for readability
\resizebox{\linewidth}{!}{%
\begin{tabular}{l c c c c c c}
\toprule
\textbf{Model} & \makecell{\textbf{Window}\\ \textbf{Length}} & \makecell{\textbf{Hop}\\ \textbf{Length}} & \makecell{\textbf{FFT}\\ \textbf{Length}} & \makecell{\textbf{Latent}\\ \textbf{Dimension}} & \makecell{\textbf{Params}\\ \textbf{(M)}} & \textbf{GMACS} \\
\midrule
DCCRN       & 400 & 100 & 512 & 1024 & 3.67  & 11.30  \\
DDAEC       & 512 & 256 & \text{NA} & 512  & 4.82  & 18.34 \\
TaylorSENet & 320 & 160 & 320 & 322  & 5.45  & 6.43   \\

GCRN        & 320 & 160 & 320 & 1024 & 9.77  & 2.42   \\
\bottomrule
\end{tabular}%
}
\label{tab:discriminativemodels}
%\vspace{-.5em}
\end{table}

\subsubsection{DCCRN} The DCCRN~\cite{hu2020dccrn} employs a complex-valued UNet-like symmetric encoder-decoder architecture with recurrent processing in the bottleneck to estimate a complex ideal ratio mask \( \mathbf{M} \). The complex \ac{Conv} for a \ac{TF}-domain noisy input \( \Stftx = \Stftx_\textrm{r} + j\Stftx_\textrm{i} \) and  a \ac{Conv} kernel \( \Stftw = \Stftw_\textrm{r} + j\Stftw_\textrm{i} \) is defined as
\begin{equation}
\Stfty = \Stftx \Stftw = (\Stftx_\textrm{r} \Stftw_\textrm{r} - \Stftx_\textrm{i} \Stftw_\textrm{i}) + j(\Stftx_\textrm{r} \Stftw_\textrm{i} + \Stftx_\textrm{i}  \Stftw_\textrm{r}) .
%\vspace{-.1cm}
\end{equation}
The encoder, consisting of five such complex-valued \ac{Conv} layers, outputs \( \Hout = \Hout_\textrm{r} + j\Hout_\textrm{i} \), which is processed by a complex-valued LSTM and a linear projection:
\begin{equation}
\begin{aligned}
\relax[\Hout'_\textrm{r}, \Hout'_\textrm{i}] &= [\LSTM_\textrm{r}(\Hout_\textrm{r}, \Hout_\textrm{i}),\ \LSTM_\textrm{i}(\Hout_\textrm{r}, \Hout_\textrm{i})] \\
\Dl &= \Stftw  \concat{\Hout'_\textrm{r}, \Hout'_\textrm{i}} + \mathbf{b} \in \RealMat{T}{1024},
\end{aligned}
\label{eq:lstm-projection}
\end{equation}
with learnable parameters \( \Stftw \) and \( \mathbf{b} \) and the conditioning information \( \Dl \in  \RealMat{T}{1024} \).

\subsubsection{DDAEC} DDAEC~\cite{pandey2020densely} is a fully convolutional, symmetric time-domain encoder-decoder model for real-time \ac{SE}. It uses dilated \ac{Conv} and densely connected blocks to capture long-range temporal dependencies. The input signal \( \vecx \) is divided into overlapping frames and encoded into a latent representation \( \Dl \in \RealMat{T}{512}\), which is used as the conditioning information. The decoder reconstructs the enhanced signal using sub-pixel convolutions and skip connections, followed by frame-wise overlap-and-add. 

\subsubsection{TaylorSENet}
\label{taylorsenet}

TaylorSENet~\cite{ijcai2022p582} approaches \ac{SE} as a Taylor series expansion with two stages: a zeroth-order approximation and a higher-order refinement. The coarse estimate \( \hat{\mathbf{S}}_0 \) is obtained using a UNet-style network \( \DNN_\psi \) and the noisy phase \( \theta_{\mathbf{X}} \):
\begin{equation}
    \hat{\mathbf{S}}_0 = \DNN_\psi(|\mathbf{X}|) \odot |\mathbf{X}| \odot  e^{j \theta_{\mathbf{X}}}.
\end{equation}
In our work, we use \( \hat{\mathbf{S}}_0 \) as the discriminative conditioning information \( \mathbf{D}_l \in \mathbb{R}^{T \times 322}\), obtained by reshaping the frequency and feature dimensions.

%
%
% Refinement is achieved by modeling the residual spectrum using higher-order derivatives
%
% \vspace{-.2cm}
% \begin{equation}
%     \hat{\mathbf{S}} = \hat{\mathbf{S}}_0 + \sum_{n=1}^{N} \frac{1}{n!} \mathcal{D}_n(\mathbf{X}, \mathbf{R}_{n-1}),
% \vspace{-.1cm}
% \end{equation}
%
%
% where \( \mathbf{R}_{n-1} \) is the residual estimated by the previous module \( \mathcal{D}_{n-1} \). 

\subsubsection{GCRN}
The gated convolutional recurrent network (GCRN)~\cite{tan2019learning} learns a complex-valued spectral mapping using an encoder-decoder architecture with $\LSTM$s in the bottleneck. The encoder comprises five \ac{Conv} layers, and two decoders estimate the real and imaginary parts of the spectrogram separately, treating them as related subtasks. Given the \ac{TF}-domain input features \( \Stftx_\textrm{r}, \Stftx_\textrm{i} \), the encoder downsamples the input via \ac{Conv} blocks, and the decoder upsamples the latent features via transposed \ac{Conv} blocks to produce \( \hatStft_\textrm{r}, \hatStft_\textrm{i} \). Both blocks utilize gated linear units ($\mathrm{GLU}$s)~\cite{van2016conditional}:
\begin{equation}
    \mathbf{Y} = \tanh(\Stftx \Stftw_1 + \mathbf{b}_1) \odot \Sigmoid(\Stftx  \Stftw_2 + \mathbf{b}_2),
%\vspace{-.1cm}
\end{equation}
where \( \Stftw_1, \Stftw_2 \) are \ac{Conv} kernels, \( \mathbf{b}_1, \mathbf{b}_2 \) are biases. GLUs enhance information flow, enabling GCRN to outperform traditional spectral mapping and masking-based methods ~\cite{tan2019learning}. The discriminative latent conditioning information is extracted via
\begin{equation}
    \Dl = \LSTM(\mathrm{ConvGLUEncoder}(\Stftx_\textrm{r}, \Stftx_\textrm{i})) \in \RealMat{T}{1024}.
%\vspace{-.1cm}
\end{equation}

\subsection{Discriminator Architecture}
We adopt a multi-scale \ac{STFT} discriminator following~\cite{defossez2022high}. It comprises multiple sub-networks with identical architectures, each operating on real and imaginary parts of the STFT representation of the input concatenated along the channel axis at different time--frequency resolutions. We employ three STFT configurations with window lengths of $2048$, $1024$, and $512$. Each sub-network starts with a 2D \ac{Conv} layer with kernel size \( (3 \times 8) \) and 32 channels, followed by three 2D \ac{Conv} layers with dilation rates of 1, 2, and 4 (along time) and a frequency stride of 2. A final 2D \ac{Conv} layer with kernel size \( (3 \times 3) \) and stride \( (1, 1) \) produces the output. LeakyReLU activations and weight normalization are applied throughout, following~\cite{du2023funcodecfundamentalreproducibleintegrable}.

\section{Experimental Details}
\label{ED}
\subsection{Training Datasets}
\subsubsection{Low- and High-SNR Datasets}
\label{traindata}

We created training datasets using the Interspeech 2020 DNS Challenge dataset \cite{reddy2020interspeech} by combining clean speech and noise at random \ac{SNR}s: $[-25, 0]$ dB for a low-\ac{SNR} dataset and $[-5, 30]$ dB for a high-\ac{SNR} dataset, both comprising approximately 1000 hours of audio. To introduce reverberation, $50\%$ of the clean utterances were convolved with randomly selected \ac{RIR}s from the Interspeech RIR dataset \cite{reddy2020interspeech}, which contains approximately 100k measured and synthetic \ac{RIR}s with reverberation times between $0.3$ and $1.5$~s.
\subsubsection{VB-DMD Dataset}
The VoiceBank-DEMAND (VB-DMD) dataset~\cite{valentini2016investigating} contains clean utterances mixed with items from the DEMAND~\cite{thiemann2013diverse} noise set, including babble, and speech-shaped noise at \ac{SNR} $\in \{0, 5, 10, 15\}$~dB.

\subsection{Evaluation Datasets}
\subsubsection{Low-SNR Dataset}
To assess performance under extremely low \ac{SNR} conditions, we curated a dataset of $1200$ samples, each $10$~s long: To this end, we mixed clean items from the synthetic, non-reverberant DNS Challenge test set~\cite{reddy2020interspeech} containing $150$ clean utterances with noise at \ac{SNR} $\in [-15, 0]$~dB. For each utterance, $8$ noise samples were randomly selected from the ESC-50 dataset~\cite{piczak2015dataset}, using a curated set of $20$ distinct noise types: \textit{rain, sea waves, crackling fire, crickets, wind, crow, washing machine, vacuum cleaner, hand saw, engine, dog, fireworks, car horn, keyboard typing, door knock, glass breaking, chainsaw, siren, helicopter}, and \textit{thunderstorm}. The dataset was grouped into four \ac{SNR} groups: $[-15, -12]$,$[-11, -8]$, $[-7, -4]$, and $[-3, 0]$~dB.

\addtocounter{footnote}{1}%
\footnotetext{The code for creating low-SNR evaluation datasets is available at this \href{https://github.com/fhgainr/Low-SNR-Eval-Data}{GitHub repository}.}

\subsubsection{VB-DMD Dataset}
The VB-DMD test set~\cite{valentini2016investigating} used utterances from two unseen speakers (“p226” and “p287”), mixed with DEMAND, babble, and speech-shaped noise at \ac{SNR} $\in \{2.5, 7.5, 12.5, 17.5\}$~dB.

\subsubsection{DNS Non-Reverb Test Dataset}
%The DNS Challenge non-reverberant test set, released with Interspeech $2020$~\cite{reddy2020interspeech}, contained clean speech from the Graz University corpus~\cite{pirker2011pitch}. 
Noisy samples were created by mixing unseen clean utterances from the Graz University corpus~\cite{pirker2011pitch} (contained in the DNS Challenge non-reverberant test set~\cite{reddy2020interspeech}) with noises from $12$ VoIP-relevant categories, at \ac{SNR} $\in[0, 25]$~dB.

\subsubsection{DNS Real Recordings}
As part of the DNS Challenge~\cite{reddy2020interspeech}, 300 noisy speech samples have been collected via Amazon Mechanical Turk using various devices and environments, covering a wide \ac{SNR} range but lacking clean references.

\subsection{Objective Metrics}
\label{OM}
To evaluate the proposed method and compare it with baseline methods, we used widely adopted intrusive and non-intrusive objective metrics.

\subsubsection{Intrusive Metrics} 
When clean references were available, we used several intrusive (full-reference) metrics: The perceptual evaluation of speech quality (PESQ) \cite{rix2001perceptual} provides scores from $1$ (poor) to $4.5$ (excellent), including narrowband and wideband variants. The frequency-weighted segmental SNR (FwSegSNR) \cite{ma2009objective} measures segmental SNR with perceptual weighting in the frequency domain. The scale-invariant signal-to-distortion ratio (SI-SDR) \cite{le2019sdr} calculates overall quality independent of the amplitude scaling. We also reported \ac{WER} and \ac{CER} using Whisper ASR \cite{radford2023robust} and JiWER \cite{morris2004and}, where reference transcriptions were generated from the clean speech signals. 

\subsubsection{Non-Intrusive Metrics} 
In cases where clean references were unavailable (e.g., real recordings), we employed non-intrusive metrics: DNSMOS \cite{reddy2021dnsmos} is a DNN-based model trained to predict mean opinion scores (MOS) based on human ratings collected via ITU-T P.808 compliant evaluations \cite{naderi2020open}. The extended DNSMOS P.835 model \cite{reddy2022dnsmos} provides scores across speech quality (SIG), background noise quality (BAK), and overall quality (OVRL), following ITU-T P.835 \cite{rec2003p}. Finally, SCOREQ \cite{ragano2024scoreq} estimates speech quality using contrastive learning, offering both non-intrusive and full-reference variants.

\subsection{Listening Test}
Objective metrics often do not correlate with human perception, especially in low \ac{SNR} conditions where assessing speech quality, intelligibility, and noise reduction is challenging \cite{objective_hu}. To address this, we conducted a multi-stimuli listening test \cite{series2014method} with fourteen participants via the webMUSHRA framework \cite{schoeffler2018webmushra}. Participants rated the overall quality of twelve randomly selected examples from the low \ac{SNR} test set, each enhanced by different algorithms. The clean signal was the reference, and the unprocessed noisy signal was included as an anchor. Scores were reported on a $0–100$ scale.

\begin{figure*}[!t]
\centering
%\hspace{-0.2cm}
\input{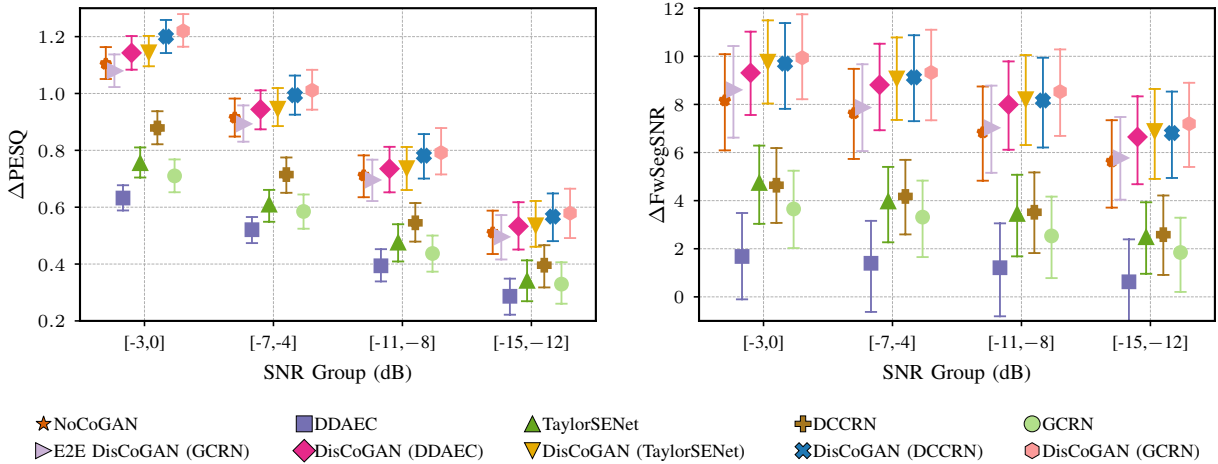}
\vspace{-0.5em}
\caption{PESQ and FwSegSNR improvements of DisCoGAN models on the low \ac{SNR} evaluation dataset using various discriminative conditioning models. The model name in parentheses indicates the discriminative model used for extracting latent features for conditioning the GAN model.}
\label{fig:discoandconditioners}
%\vspace{-1em}
\end{figure*}

\subsection{DisCoGAN Training Details}
\label{DisCoGAN TD}
We trained the models at a $16$~kHz sampling rate in two steps. First, we trained the discriminative models (Sec.~\ref{Discriminative Models}), namely DDAEC \cite{pandey2020densely}, GCRN \cite{tan2019learning}, DCCRN \cite{hu2020dccrn}, and TaylorSENet \cite{ijcai2022p582} using the low \ac{SNR} training dataset. All models were trained with their original configurations as specified in the respective papers. It is important to note that all of these models used different window lengths and hop sizes, resulting in varying time resolutions as described in Table~\ref{tab:discriminativemodels}.

In the next step, we trained DisCoGAN while extracting latent features $\Dl$ from the discriminative model with frozen parameters as mentioned in Sec.~\ref{Discriminative Models}. To match the time dimension of $\Dl$, with the generative latent features 
 \(\Gl\), as mentioned in Sec.~\ref{DiscCond}, linear interpolation \cite{chapra2011numerical} was performed.
 We followed similar training procedures as outlined in \cite{du2023funcodecfundamentalreproducibleintegrable} for training the generator and discriminator of the GAN model. Training was performed on a Tesla-A100 GPU with a batch size of $16$ for $600$~k iterations. To avoid dominance of the discriminator, it was updated only when its loss exceeded that of the generator. The hyperparameters $\lambda_\text{t}$, $\lambda_\text{f}$, $\lambda_\text{adv}$, and $\lambda_\text{feat}$ were set to $1.0$, $1.0$, $\frac{1}{9}$, and $\frac{100}{9}$, respectively. For the generator \ac{STFT} parameters, we used an FFT length of $512$, a window size of $512$, and a hop length of $160$.

\subsection{Discussed Models}
We compared our proposed approach against a set of generative and discriminative baselines. Many of these baselines were trained from scratch, while for others, we either employed publicly available pre-trained models or reported the objective metrics as presented in the original publications for the respective datasets.

\subsubsection{Proposed DisCoGAN Models}
We trained the DisCoGAN model with four different discriminative conditioning models (Sec.~\ref{Discriminative Models}): DisCoGAN (GCRN), DisCoGAN (DCCRN), DisCoGAN (DDAEC), and DisCoGAN (TaylorSENet), where the model name in parentheses indicates the discriminative conditioning model. Unless specified otherwise, DisCoGAN (GCRN) is used as the default variant throughout the paper.

\subsubsection{E2E GAN Models} For \ac{E2E} GAN-based baselines, we trained NoCoGAN and SEANet, E2E DisCoGAN (GCRN). NoCoGAN shares the same architecture as DisCoGAN, but without discriminative conditioning (Sec.~\ref{DA}) and was trained using a similar procedure, detailed in Sec.~\ref{DisCoGAN TD}. The E2E DisCoGAN (GCRN) shares the same configuration as the proposed DisCoGAN (GCRN). However, in this case, the encoder of the GCRN model was trained from scratch alongside the generator of the DisCoGAN model, using the same learning objective described in Section~\ref{LO}.  Additionally, we trained SEANet \cite{tagliasacchi2020seanetmultimodalspeechenhancement} with an additional  bottleneck LSTM. We also included HiFi-GAN \cite{su2020hifi}, SEGAN \cite{pascual2017segan}, and MetricGAN+ \cite{fu2021metricgan+} as representative generative baselines from the literature.

\subsubsection{ Two Stage GANs}
For GAN-last-based baselines, where a generative model follows a discriminative \ac{SE} stage, we trained GCRN+NoCoGAN and GCRN+DisCoGAN. For the GAN-first approaches, where the generative model is followed by a discriminative model, we trained NoCoGAN+GCRN and DisCoGAN+GCRN. In both configurations, the second-stage model is trained on the output of the first-stage model with frozen weights.

\subsubsection{Discriminatives Methods}
We also trained multiple discriminative baseline models, including all conditioning models: GCRN, DCCRN, DDAEC, and TaylorSENet. Additionally, we trained NoCoGAN-D and DisCoGAN-D, sharing the same architectures as NoCoGAN and DisCoGAN, respectively, but are trained solely with the reconstruction loss $\mathcal{L}_{\text{rec}}$, without adversarial losses.

\subsubsection{Additional Generative Model Baselines}
The diffusion-based models SGMSE \cite{welker2022speech}, SGMSE+ \cite{richter2023speech}, CDiffuse \cite{lu2022conditional}, and the VAE-based RVAE \cite{leglaive2020recurrent} were also included as baselines.

\begin{table*}[htbp]
\caption{Evaluation of the proposed DisCoGAN method against other GAN-based \ac{SE} approaches in extremely low SNR scenarios. The result of the best-performing model per SNR group is denoted in bold.}
\small
\centering
\resizebox{\textwidth}{!}{%
\begin{tabular}{l l c c c c c c c c c c c c}
    \toprule
    \multicolumn{2}{l}{} & \multicolumn{4}{c}{\textbf{$\Delta$PESQ (↑)}} & \multicolumn{4}{c}{\makecell{\textbf{SI-SDR (↑)} \\\textbf{(dB)}}} & \multicolumn{4}{c}{\textbf{$\Delta$FwSegSNR (↑)}} \\ 
    \cmidrule(lr){3-6} \cmidrule(lr){7-10} \cmidrule(lr){11-14}
    \textbf{Method} & \textbf{Model} & \makecell{[-15,-12]\\dB} & \makecell{[-11,-8]\\dB} & \makecell{[-7,-4]\\dB} & \makecell{[-3,0]\\dB} & \makecell{[-15,-12]\\dB} & \makecell{[-11,-8]\\dB} & \makecell{[-7,-4]\\dB} & \makecell{[-3,0]\\dB} & \makecell{[-15,-12]\\dB} & \makecell{[-11,-8]\\dB} & \makecell{[-7,-4]\\dB} & \makecell{[-3,0]\\dB} \\
    \midrule
    \multirow{2}{*}{\ac{E2E}}& NoCoGAN & 0.51 & 0.71 & 0.91 & 1.10 & 16.85 & 15.51 & 14.20 & 12.42 & 5.63 & 6.84 & 7.63 & 8.16 \\
    & \ac{E2E} DisCoGAN (GCRN) & 0.50 & 0.70 & 0.89 & 1.08 & 16.86 & 15.46 & 14.06 & 12.25 & 5.77& 7.03 & 7.87 & 8.61 \\
    \midrule
        \multirow{2}{*}{GAN-first} 
        & NoCoGAN+GCRN & 0.49 & 0.68 & 0.89 & 1.07 & 16.94 & 15.56 & 14.23 & 12.39 & 4.95 & 6.17 & 7.01 & 7.46 \\
        & DisCoGAN+GCRN & 0.56 & 0.76 & 0.99 & 1.18 & \textbf{17.53} & \textbf{16.06} & 14.78 & 12.91 & 6.86 & 8.10 & 8.92 & 9.49 \\
          \midrule
    \multirow{2}{*}{GAN-last} & GCRN+NoCoGAN & 0.41 & 0.58 & 0.75 & 0.92 & 15.73 & 14.71& 13.39 & 11.36 & 3.27 & 4.23 & 4.89 & 4.97 \\
     &GCRN+DisCoGAN & 0.47 & 0.64 & 0.85 & 1.05 & 16.12 & 15.27 & 13.99 & 12.27 & 4.46 & 5.86 & 6.64 & 7.22 \\ 
    \midrule
    \textbf{Proposed DisCoGAN} & DisCoGAN (GCRN) & \textbf{0.58} & \textbf{0.79} & \textbf{1.01} & \textbf{1.22} & 17.48 & \textbf{16.06} & \textbf{14.81} & \textbf{12.96} & \textbf{7.19} & \textbf{8.53} & \textbf{9.33} & \textbf{9.94} \\
    \bottomrule
\end{tabular}
}
\label{tab:commonganapproaches}
\end{table*}

\section{Results}
\label{results}

We first evaluate DisCoGAN using various pretrained discriminative conditioning models (Sec.~\ref{DisCoGAN+CM}). We then compare DisCoGAN with existing GAN-based approaches, including \ac{E2E}, GAN-first and GAN-last methods in low \ac{SNR} scenarios (Sec.~\ref{DisCoGAN+Comp}). Finally, we assess DisCoGAN's performance across diverse \ac{SNR} conditions and real-world data (Secs.~\ref{SNRSDisCo} and \ref{RealData}), and present ablation results focusing on sensitivity and temporal dependencies of the conditioning discriminative latent features (Sec.~\ref{AS}).

\subsection{DisCoGAN with Diverse Conditioning Models} 
\label{DisCoGAN+CM}
Figure~\ref{fig:discoandconditioners} shows the PESQ and FwSegSNR improvements of DisCoGAN conditioned with different discriminative models. All DisCoGAN variants outperform the baseline discriminative models, NoCoGAN and E2E DisCoGAN (GCRN). Among them, DisCoGAN (GCRN) achieves the best results with mean PESQ and FwSegSNR improvements of $1.22$ and $9.94$ dB, respectively, in the $[0, -3]$ dB SNR group. Similar trends are observed in other SNR groups. The DCCRN model outperforms other discriminative models with PESQ improvements of $0.88$, $0.71$, $0.55$, and $0.40$ across decreasing SNR intervals. TaylorSENet and GCRN follow in performance. However, DisCoGAN (GCRN) showed the best performance even though GCRN was inferior relative to DCCRN when used directly for \ac{SE}. DisCoGAN (DCCRN) is second in performance, while DisCoGAN (TaylorSENet) and DisCoGAN (DDAEC) show a slightly lower performance. An important point to note is that E2E DisCoGAN (GCRN) performs comparably to NoCoGAN, achieving slightly lower PESQ but better FwSegSNR improvements. This suggests that the performance gains observed in the proposed DisCoGAN variants are not due to the added complexity of using an additional feature encoder. Instead, the improvement can be attributed to the use of latent features from a pretrained discriminative SE model to condition the GAN.

\begin{figure}[htbp]
\centering
%\vspace{-1em}
%\hspace{-0.4cm}
  % Adjust width as needed:
    \input{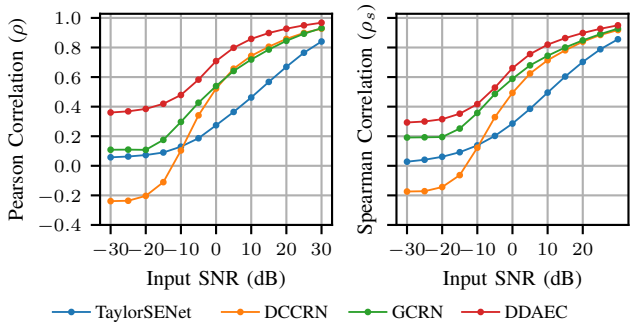}
\vspace{-1em}
\caption{Pearson \(\rho\) and Spearman \(\rho_s\) correlation coefficients of the latent features at various SNR levels.}
\label{fig:latentcorrelation}
%\vspace{-1.5em}
\end{figure}

To understand the performance differences among the DisCoGAN variants, we analyze the Pearson ($\rho$) and Spearman ($\rho_s$) correlation coefficients between latent features extracted at various SNR levels and those from extraced clean speech. For this we first process the clean speech signal and its noisy counterparts at different SNR levels using the different discriminative conditioning models (Sec.~\ref{Discriminative Models}). Then, we compute how strongly the latent features at each SNR level correlate with those from the clean speech. We hypothesize that effective discriminative conditioning features should show low correlation at low SNR and high correlation at high SNR to properly represent the SNR and various noise conditions. As shown in Fig.~\ref{fig:latentcorrelation}, the DDAEC model achieves the highest correlation at high SNRs and maintains strong correlation even at low SNRs. In contrast, TaylorSENet shows low correlation at $\text{SNR} = -30$\,dB ($\rho = 0.06$, $\rho_\textrm{s} = 0.02$) and high correlation at $\text{SNR} = 30$\,dB ($\rho = 0.84$, $\rho_\textrm{s} = 0.86$), though still lower than DCCRN, GCRN, and DDAEC ($\rho = 0.93$, $0.93$, and $0.97$, respectively, at 30\,dB). This can be explained by its architecture (Sec.~\ref{taylorsenet}), which includes a degraded noisy phase component, which limits high-SNRs performance. GCRN and DCCRN follow the hypothesized trends, with DCCRN even showing a negative correlation at low SNRs.  This characteristic likely contributes to the superior performance observed in the DisCoGAN variants DisCoGAN (GCRN) and DisCoGAN (DCCRN) that utilize the discriminative latent features from GCRN and DCCRN. In the rest of the paper, we use \ac{DisCoGAN} (GCRN) as the default \ac{DisCoGAN} variant for comparison, as it achieves the best performance among all variants.

\begin{figure*}[htbp]
\centering
%\hspace{-0.2cm}
\input{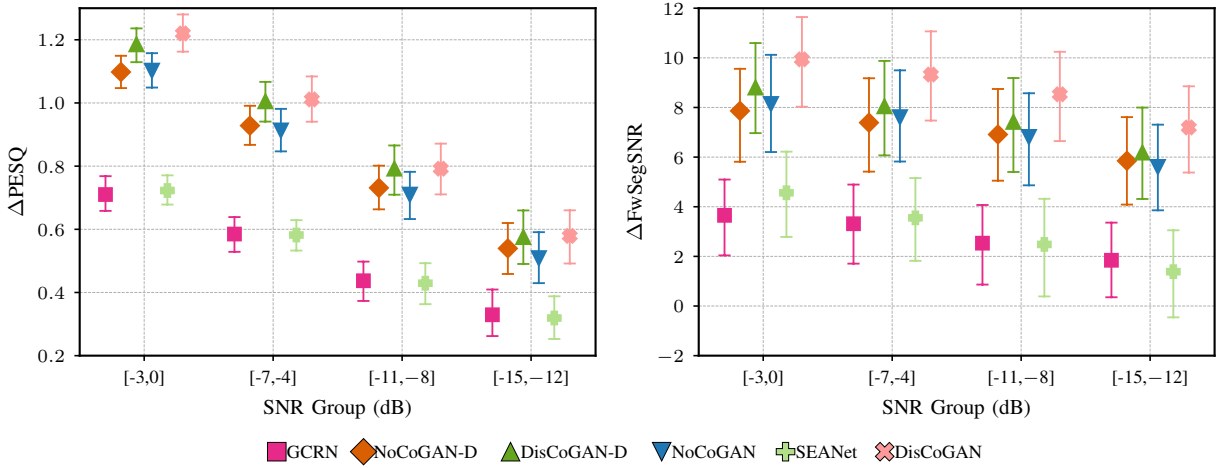}
\vspace{-1em}
\caption{PESQ and FwSegSNR improvement for the DisCoGAN vs baseline generative and discriminative models in low SNR scenarios.}
\label{fig:lowsnrbox}
%\vspace{-1em}
\end{figure*}

\begin{figure}[htbp]
  \centering
  \makebox[\textwidth][l]{%
    \hspace{0.2cm}%
    \input{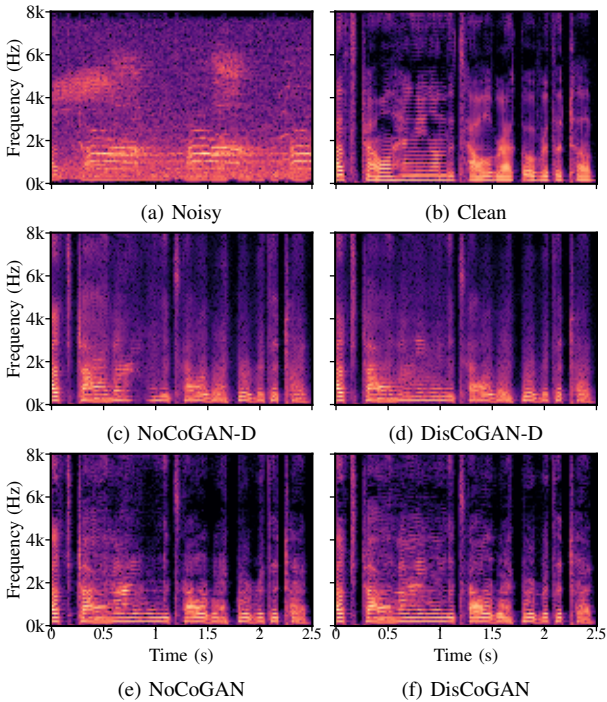}%
  }
  \vspace{-2em}
  \caption{Spectrograms of (a) a noisy speech signal corrupted by strong noise at -8 dB SNR from (b) a clean speech signal, enhanced by: (c) NoCoGAN-D (d) DisCoGAN-D (e) NoCoGAN and (f) DisCoGAN.}
  %\vspace{-2em}
  \label{figure:spectorgram}
\end{figure}

\addtocounter{footnote}{1}%
\footnotetext{Demo audio samples are provided at this \href{https://fhgainr.github.io/DisCoGAN-Journal/}{ demo site}.}

\begin{table*}[!t]
\caption{Evaluation of DisCoGAN and generative and discriminative baselines using non-intrusive and intrusive metrics for low SNRs.}
%\vspace{-0.1cm}
\small
\centering
\resizebox{\textwidth}{!}{%
\begin{tabular}{llcccccccccccc}
    \toprule
    \multicolumn{2}{l}{} 
    & \multicolumn{4}{c}{\textbf{\makecell{SCOREQ\\MOS (↑)}}} 
    & \multicolumn{4}{c}{\textbf{\makecell{SCOREQ\\Distance (↓)}}} 
    & \multicolumn{4}{c}{\textbf{\makecell{DNSMOS\\P808\_MOS (↑)}}}\\
    \cmidrule(lr){3-6} \cmidrule(lr){7-10} \cmidrule(lr){11-14}
    \textbf{Method} & \textbf{Model} & \makecell{[-15,-12]\\dB} & \makecell{[-11,-8]\\dB} & \makecell{[-7,-4]\\dB} & \makecell{[-3,0]\\dB} 
                   & \makecell{[-15,-12]\\dB} & \makecell{[-11,-8]\\dB} & \makecell{[-7,-4]\\dB} & \makecell{[-3,0]\\dB} 
                   & \makecell{[-15,-12]\\dB} & \makecell{[-11,-8]\\dB} & \makecell{[-7,-4]\\dB} & \makecell{[-3,0]\\dB} \\
    \midrule
    \multirow{2}{*}{Unprocessed} 
        & Noisy & 1.50 & 1.65 & 1.81 & 2.04 & 1.36 & 1.31 & 1.27 & 1.19 & 2.52 & 2.63 & 2.73 & 2.83 \\
        & Clean & 4.57 & 4.59 & 4.59 & 4.59 & 0 & 0 & 0 & 0 & 4.00 & 4.02 & 4.01 & 4.03 \\
    \midrule
    \multirow{3}{*}{Discriminative} 
        & GCRN & 1.95 & 2.25 & 2.60 & 2.98 & 1.16 & 1.06 & 0.94 & 0.79 & 3.03 & 3.26 & 3.44 & 3.64 \\
        & NoCoGAN-D & 2.30 & 2.67 & 3.01 & 3.32 & 1.05 & 0.91 & 0.77 & 0.63 & 3.18& 3.41 & 3.54 & 3.71 \\
        & DisCoGAN-D & 2.39 & 2.76 & 3.13 & 3.47 & 1.01 & 0.87 & 0.72 & 0.56 & 3.23 & 3.47 & 3.62 & 3.79 \\
    \midrule
    \multirow{2}{*}{Generative} 
        & NoCoGAN & 2.29 & 2.67 & 3.04 & 3.39 & 1.04 & 0.91 & 0.77 & 0.61 & 3.70 & 3.83 & 3.90 & 3.98 \\
        & \textbf{DisCoGAN} & \textbf{2.51} & \textbf{2.92} & \textbf{3.34} & \textbf{3.69} & \textbf{0.97} & \textbf{0.81} & \textbf{0.65} & \textbf{0.49} & \textbf{3.84} & \textbf{3.95} & \textbf{4.00} & \textbf{4.06} \\
    \bottomrule
\end{tabular}
}
%\vspace{-.2cm}
\label{tab:scoreqandDNSMos}
\end{table*}

\subsection{Comparison with Existing GAN-Based Approaches}
\label{DisCoGAN+Comp}

In Table~\ref{tab:commonganapproaches}, we report the performance of DisCoGAN compared to GAN-based \ac{SE} baselines, including \ac{E2E} and two-stage GANs on the low-SNR dataset. DisCoGAN outperforms all other methods for all objective metrics and SNR groups. In particular, in terms of $\Delta \text{FwSegSNR}$, DisCoGAN shows at least a $2.5$\,dB gain over the best GAN-last models, and an average $0.4$\,dB improvement over the best performing GAN-first model in all SNR groups. GAN-last methods show the worst performance for all metrics. This aligns with the poor performance of the GCRN model shown in Fig.~\ref{fig:discoandconditioners}, where, for extremely low SNRs, we observed, the discriminative GCRN model tends to suppress speech components that cannot be recovered by the subsequent generative model. However, GCRN+DisCoGAN outperforms GCRN+NoCoGAN, as DisCoGAN takes advantage of both noisy latent features and estimated clean speech from the preceding discriminative model, showing the effectiveness of  the proposed discriminative conditioning technique. The GAN-first DisCoGAN+GCRN model performs similarly to DisCoGAN alone but shows slight drops in PESQ and FwSegSNR, likely because the discriminative model suppresses noise but may also remove parts of the generated content it deems noisy, improving SI-SDR while slightly degrading other metrics.

\subsection{Performance Across Various SNR Scenarios}
\label{SNRSDisCo}
\subsubsection{Low-SNR Scenarios}
\label{sec:lowsnrscenarios}
We report the mean $\Delta \text{PESQ}$ and $\Delta \text{FwSegSNR}$ for DisCoGAN and the baselines for the low-SNR dataset in Fig.~\ref{fig:lowsnrbox}. DisCoGAN consistently outperforms all baselines for both metrics and SNR groups. Interestingly, the discriminatively trained models NoCoGAN-D and DisCoGAN-D achieve objective scores comparable with their GAN-based counterparts. However, by inspecting the spectrograms in Fig.~\ref{figure:spectorgram}, we see that discriminative models often produce overly smooth high-frequency components when speech is heavily masked by noise. In contrast, GAN-based models generate fine-grained, speech-like high-frequency structures overlooked by traditional objective metrics, suggesting an inherent bias of these metrics toward discriminative \ac{SE} methods.

We further evaluate these methods using the DNN-based intrusive and non-intrusive metrics, SCOREQ and DNSMOS, in Tab.~\ref{tab:scoreqandDNSMos}. DisCoGAN outperforms all baselines for both metrics, with a clear margin over its discriminative counterpart, DisCoGAN-D. Although SCOREQ results are similar for NoCoGAN and NoCoGAN-D, DNSMOS shows a stronger advantage for GAN-based models. In particular, at higher SNRs, e.g., $[-7,0]$ dB, DisCoGAN sometimes scores equal to or above the clean reference, highlighting DisCoGAN's generative capabilities and possible leniency of non-intrusive metrics toward hallucinated content. Table~\ref{tab:wer_cer_comparison_new} presents \ac{WER} and \ac{CER} results, showing DisCoGAN’s effectiveness with up to $5\%$ improvement over competing methods for SNRs down to $-7$ dB. However, at very low SNRs (~$-10$ dB), where speech is almost unintelligible, ASR performance degrades sharply, with a WER exceeding $200\%$ on noisy signals. In these conditions, discriminative models outperform generative ones, likely because generative models hallucinate content, increasing recognition errors. It is also important to note that WER/CER results may include overestimated or underestimated scores, as the ASR system might either compensate for halucinated content or introduce hallucinations.

\begin{table}[t]
\caption{WER/CER (in $\%$) on low-SNR evaluation dataset.}
\centering
\begin{tabular}{lcccc}
\toprule
& \multicolumn{4}{c}{\textbf{WER (↓) $|$ CER (↓)}} \\
\cmidrule(lr){2-5}
\makecell{\textbf{Method}} & \makecell{[-15,-12]\\dB} & \makecell{[-11,-8]\\dB} & \makecell{[-7,-4]\\dB} & \makecell{[-3,0]\\dB} \\
\midrule
Noisy     & \makecell{496 $|$ 730} & \makecell{250 $|$ 239} & \makecell{90 $|$ 66}   & \makecell{39 $|$ 24} \\
GCRN         & \makecell{181 $|$ 179} & \makecell{88 $|$ 76}   & \makecell{73 $|$ 58}   & \makecell{43 $|$ 29} \\
NoCoGAN-D       & \makecell{\textbf{87 $|$ 75}}  & \makecell{72 $|$ 60}   & \makecell{51 $|$ 37}   & \makecell{29 $|$ 18} \\
DisCoGAN-D      & \makecell{91 $|$ 82}   & \makecell{\textbf{62} $|$ 54}  & \makecell{44 $|$ 28}   & \makecell{29 $|$ 17} \\
NoCoGAN         & \makecell{113 $|$ 116} & \makecell{90\hphantom{0} $|$ 105}  & \makecell{43 $|$ 28}   & \makecell{31 $|$ 19} \\
\textbf{DisCoGAN } &\makecell{104 $|$ 96\hphantom{0}}  & \makecell{68 $|$ \textbf{46}}   & \makecell{\textbf{39 $|$ 25}}   & \makecell{\textbf{26 $|$ 16}} \\

\bottomrule
\end{tabular}
\label{tab:wer_cer_comparison_new}
%\vspace{-0.2cm}
\end{table}

\begin{table}[t]
\caption{Results on the DNS challenge non-reverb dataset~\cite{reddy2020interspeech}.}
\footnotesize
\centering
\setlength{\tabcolsep}{2.5pt}
\begin{tabular}{
    l
    l
    S[table-format=1.2]
    S[table-format=1.2]
    S[table-format=2.2, detect-none]
    S[table-format=1.2, detect-none]
}
\toprule
\textbf{Method} & {\makecell{\textbf{Training} \\ \textbf{Dataset}}} & {\makecell{\textbf{SCOREQ}\\\textbf{MOS (↑)}}} & {\makecell{\textbf{DNS} \\ \textbf{MOS (↑)}}} & {\makecell{\textbf{SI-SDR (↑)} \\\textbf{(dB)}}} & {\textbf{PESQ (↑)}}\\
\midrule
Noisy & -- & 2.78 & 3.15 & 9.06 & 1.58 \\
Clean & -- & 4.60 & 4.01 & {NA} & {NA} \\
\midrule
\multirow{2}{*}{NoCoGAN}
& Low SNR & 4.10 & 4.06 & 18.13 & 3.12 \\
& High SNR & 4.17 & 4.04 & 17.82 & 3.22 \\
\midrule
\multirow{2}{*}{\textbf{DisCoGAN}}
& Low SNR & \textbf{4.32} & \textbf{4.12} & \textbf{18.83} & 3.26 \\
& High SNR & 4.27 & 4.08 & 18.74 & \textbf{3.30} \\
\bottomrule
\end{tabular}
\label{tab:DNSnoreverb}
%\vspace{-1em}
\end{table}

\begin{table}[!t]
\caption{Results on the standard VB-DMD dataset \cite{valentini2016investigating}.}
\footnotesize
\centering
\setlength{\tabcolsep}{1.4pt}
\begin{tabular}{
    l
    l
    S[table-format=2.1, detect-none]
    S[table-format=1.2]
    S[table-format=2.2]
    S[table-format=1.2]
}
\toprule
\textbf{Method} & {\makecell{\textbf{Training} \\ \textbf{Dataset}}} & {\makecell{\textbf{Params} \\ \textbf{(M)}}} & {\textbf{PESQ (↑)}} & {\makecell{\textbf{SI-SDR (↑)} \\ \textbf{(dB)}}} & {\makecell{\textbf{DNS} \\ \textbf{MOS (↑)}}} \\
\midrule
Noisy & {--} & {--} & 1.97 & 8.4 & 3.09 \\
RVAE \cite{leglaive2020recurrent} & VB-DMD & {--} & 2.43 & 16.4 & 3.30 \\
HiFiGAN \cite{su2020hifi} & VB-DMD & {--} & 2.94 & {--} & {--} \\
MetricGAN+ \cite{fu2021metricgan+} & VB-DMD & {--} & \textbf{3.13} & 8.5 & 3.37 \\
SEGAN \cite{pascual2017segan} & VB-DMD & 97.5 & 2.16 & {--} & {--} \\

SGMSE+ \cite{richter2023speech} & VB-DMD & 65 & 2.93 & 17.3 & 3.56 \\
\midrule
\multirow{3}{*}{\textbf{DisCoGAN}}
& VB-DMD &  44 & 3.08 & 18.06 & 3.51 \\
& Low SNR &  44 & 2.85 & 17.83 & 3.64 \\
& High SNR &  44 & 2.95& \textbf{18.63} & \textbf{3.65} \\
\bottomrule
\end{tabular}
\label{tab:VCTK}
%\vspace{-1em}
\end{table}

\begin{table}[htbp]
\caption{Results on the DNS real recordings test set~\cite{reddy2020interspeech}.}
\footnotesize
\centering
\setlength{\tabcolsep}{3pt}
\begin{tabular}{
    l
    S[table-format=1.2]
    S[table-format=1.2]
    S[table-format=1.2]
    S[table-format=1.2]
}
\toprule
\textbf{Method} & {\textbf{DNSMOS} (↑)} & {\textbf{SIG} (↑)} & {\textbf{BAK} (↑)} & {\textbf{OVRL} (↑)} \\
\midrule
Noisy & 3.05 & 3.05 & 2.51 & 2.26 \\
RVAE \cite{leglaive2020recurrent} & 3.29 & 3.16 & 2.91 & 2.44 \\
CDiffuse \cite{lu2022conditional} & 3.14 & 3.15 & 3.19 & 2.55 \\
MetricGAN+ \cite{fu2021metricgan+} & 3.26 & 2.88 & 3.39 & 2.45 \\
SGMSE \cite{welker22_interspeech} & 3.38 & 3.22 & 3.02 & 2.52 \\
SGMSE+ \cite{richter2023speech} & 3.64 & \textbf{3.42} & 3.82 & \textbf{3.04} \\
\textbf{DisCoGAN} & \textbf{3.65} & 3.32 & \textbf{3.91} & 2.98 \\
\bottomrule
\end{tabular}
\label{tab:DNSReal}
\end{table}

\subsubsection{High-SNR Scenarios}

To evaluate DisCoGAN’s generalization abilities, we tested it on the DNS Challenge non-reverberant dataset and trained it on low and high \ac{SNR} data (Sec.~\ref{traindata}). Table~\ref{tab:DNSnoreverb} shows that DisCoGAN consistently outperforms NoCoGAN. Models trained on low SNR data achieve better SI-SDR, while high-SNR training improves PESQ. Notably, DNSMOS rates GAN-based methods higher than the clean reference, e.g., DisCoGAN scores \(4.12\) versus \(4.01\) for clean speech, suggesting that generative outputs are more perceptually appealing to the DNN-based MOS predictor, despite slight deviations from the ground truth. Table~\ref{tab:VCTK} shows results of DisCoGAN trained on various datasets and evaluated on VB-DMD testset in comparison to other generative baselines. DisCoGAN trained on high SNRs achieves the best SI-SDR (\(18.63\,\mathrm{dB}\)) and DNSMOS (\(3.65\)) values. DisCoGAN trained on VB-DMD yields a PESQ of \(3.08\), the second-best result after MetricGAN+, which was optimized for PESQ but scores lower in SI-SDR and DNSMOS. Importantly, DisCoGAN trained on low SNR data, representing an unmatched test condition, still outperforms baselines with SI-SDR of \(17.83\,\mathrm{dB}\), DNSMOS \(3.64\), and competitive PESQ  of \(2.85\), demonstrating  strong generalization capability.

\begin{figure}[t]
  \centering
  % Scale down the entire plot to reduce height
  \resizebox{\linewidth}{0.8\linewidth}{%
    \input{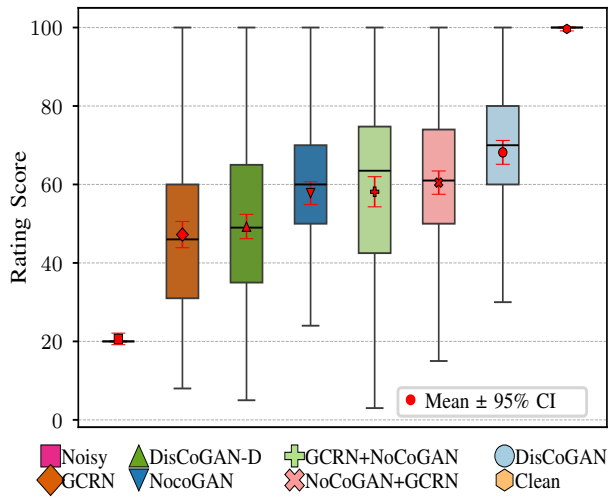}
  }
  \vspace{-1.5em}
  \caption{Results of the multi-stimuli listening test. The red circles indicate the mean rating scores along with the $95\%$ confidence interval.}
  \label{fig:lt}
  \vspace{-0.5em}
\end{figure}

\subsection{Performance on Real Data}  
\label{RealData}
Complementing experiments on simulated data, we evaluate the \ac{SE} performance of DisCoGAN and baseline generative methods on real-world noisy recordings from the DNS real recordings dataset. As shown in Tab.~\ref{tab:DNSReal}, DisCoGAN outperforms all baselines with \(3.65\) in DNSMOS and \(3.91\) in BAK (MOS), respectively. In SIG (MOS) and OVRL (MOS), DisCoGAN ranks second, behind SGMSE+ with scores of \(3.42\) and \(3.04\). Notably, SGMSE+ performs joint dereverberation and denoising, which boosts SIG and OVRL, while DisCoGAN focuses solely on noise reduction. These results highlight DisCoGAN’s robustness in real-world noisy conditions.

\subsection{Listening Test} 
\label{LT}

The listening test results, as shown in Fig.~\ref{fig:lt}, show that DisCoGAN was also preferred by human listeners, achieving the highest mean and median rating scores of $68.17$ and $70$, respectively. The baselines E2E NoCoGAN, and the two-stage NoCoGAN+GCRN and GCRN+NoCoGAN, perform similarly, with mean scores of $57.78$, $60.47$, and $58.13$. The discriminative models receive the lowest ratings: GCRN obtains a mean score of $47.22$, while DisCoGAN-D scores $49.27$. These results support our hypothesis (Sec.~\ref{sec:lowsnrscenarios}) that traditional objective metrics such as PESQ, SI-SDR, and FwSegSNR may favor discriminative models. For instance, DisCoGAN-D performs comparably to DisCoGAN on those metrics, but is rated significantly lower by human listeners. We also observe a higher variance in the rating scores of the GAN-last model (GCRN+NoCoGAN). Although it achieves a higher median rating than the GAN-first model (NoCoGAN+GCRN), its mean score is lower. As discussed in Sec.~\ref{DisCoGAN+Comp}, at extremely low SNRs, the GCRN model tends to suppress speech components that cannot be recovered by the subsequent GAN model, thereby lowering the average perceived quality.

\begin{table}[!t]
\caption{Impact of component removal on DisCoGAN performance.}
\footnotesize
\setlength{\tabcolsep}{3pt} % reduced column spacing
\centering
\begin{tabular}{l S[table-format=1.2, table-align-text-after=false] S[table-format=2.2, table-align-text-after=false] S[table-format=-1.2, table-align-text-after=false]}
\toprule
& {\textbf{$\Delta$PESQ (↑)}} & {\makecell{\textbf{$\Delta$SI-SDR (↑)}\\\textbf{(dB)}}} & {\makecell{\textbf{$\Delta$FwSegSNR (↑)}\\\textbf{(dB)}}} \\
\midrule
DisCoGAN & 1.22 & 12.74 & 9.94 \\
\midrule
-- Disc. Cond. & 1.10\textsubscript{\scriptsize{-9.8\%}} & 12.21\textsubscript{\scriptsize{-4.2\%}} & 8.16\textsubscript{\scriptsize{-17.9\%}} \\
-- FiLM & 1.07\textsubscript{\scriptsize{-12.3\%}} & 12.04\textsubscript{\scriptsize{-5.5\%}} & 7.61\textsubscript{\scriptsize{-23.4\%}} \\
-- Skip Conns. & 0.82\textsubscript{\scriptsize{-32.8\%}} & 3.28\textsubscript{\scriptsize{-74.3\%}} & -0.27\textsubscript{\scriptsize{-102.7\%}} \\
\bottomrule
\end{tabular}
\label{tab:Components}
\end{table}

\begin{table}[htbp!]
\caption{Mean $\Delta$PESQ and $\Delta$SI-SDR improvements for DisCoGAN across different conditioning SNR levels (rows) and input SNR ranges
(columns).}
\centering
\footnotesize
\setlength{\tabcolsep}{3pt} % Reduced column separation
\begin{tabular}{l *{3}{S[table-format=-1.2]} *{3}{S[table-format=-2.2]}}
\toprule
\multirow{2}{*}{\textbf{Input SNR}} & \multicolumn{3}{c}{\textbf{$\Delta$PESQ (↑)}} & \multicolumn{3}{c}{\textbf{$\Delta$SI-SDR (dB) (↑)}} \\
\cmidrule(lr){2-4} \cmidrule(lr){5-7}
& {\makecell{[-11,-8]\\dB}} & {\makecell{[-7,-4]\\dB}} & {\makecell{[-3,0]\\dB}} & {\makecell{[-11,-8]\\dB}} & {\makecell{[-7,-4]\\dB}} & {\makecell{[-3,0]\\dB}} \\
\midrule
Clean & 0.47 & 0.62 & 0.76 & 11.85 & 11.27 & 9.88 \\
30dB & 0.56 & 0.79 & 1.01 & 13.54 & 13.20 & 11.86 \\
25dB & 0.60 & 0.83 & 1.07 & 13.97 & 13.52 & 12.21 \\
20dB & 0.65 & 0.89 & 1.14 & 14.46 & 13.76 & 12.52 \\
15dB & 0.70 & 0.94 & 1.19 & 14.92 & 14.06 & 12.76 \\
10dB & 0.74 & 0.98 & 1.22 & 15.35 & 14.44 & 12.91 \\
5dB & 0.77 & 1.00 & 1.22 & 15.73 & 14.70 & \textbf{\hphantom{0-}13.05} \\
\midrule
\multicolumn{1}{l}{\makecell[l]{Matching\\(Noisy)}} & \textbf{\hphantom{0-}0.79} & \textbf{\hphantom{0-}1.01} & \textbf{\hphantom{0-}1.22} & 16.06 & 14.81 & 12.96 \\
\midrule
-5dB & 0.78 & 0.99 & 1.18 & \textbf{\hphantom{0-}16.18} & \textbf{\hphantom{0-}14.91} & 13.00 \\
-10dB & 0.76 & 0.96 & 1.13 & 16.18 & 14.85 & 12.80 \\
-15dB & 0.74 & 0.93 & 1.06 & 16.04 & 14.57 & 12.05 \\
-20dB & 0.70 & 0.85 & 0.86 & 15.57 & 13.25 & 9.43 \\
-25dB & 0.64 & 0.69 & 0.56 & 14.77 & 11.14 & 4.79 \\
-30dB & 0.52 & 0.46 & 0.31 & 12.99 & 7.66 & 1.12 \\
\multicolumn{1}{l}{Noise} & 0.05 & 0.02 & -0.05 & 7.00 & 2.40 & -2.74 \\
\bottomrule
\end{tabular}
\label{tab:sensitivitynonmatchingsignal}
\end{table}

\subsection{Ablation Study}
\label{AS}

\subsubsection{Component-wise Ablation of DisCoGAN}

The ablation study in Table~\ref{tab:Components} shows that removing discriminative conditioning significantly reduces performance for all metrics, i.e., $\Delta$PESQ is reduced by 9.8\%, $\Delta$SI-SDR by 4.2\%, and $\Delta$FwSegSNR by 17.9\%. While removing the \ac{FiLM} conditioning in skip connections further degraded performance, the largest declines occurred when both components were removed, causing drops of 32.8\%, 74.3\%, and 102.7\%, respectively.

\subsubsection{Sensitivity to Non-Matching Latent Features}

Here, we evaluated DisCoGAN’s performance when the discriminative conditioning model received either matching or non-matching inputs relative to the generator input during inference. The generator always received the original noisy signal, while the conditioning model was fed signals at various SNRs $\in[-30,30]$\,dB, including clean, noise-only, and the matching noisy signal.
Table~\ref{tab:sensitivitynonmatchingsignal} shows stable performance when the conditioning signal's \ac{SNR} was close to that of the generator input, with a bias toward higher \ac{SNR}s, e.g., in the $[-3, 0]$\,dB group, PESQ improvement remained unchanged even when the input to the conditioning model was at $5$ or $10$\,dB. SI-SDR improved slightly when the input to the conditioning model had matched or slightly lower \ac{SNR} than the input, e.g., in the $[-11, -8]$\,dB group, SI-SDR increased by around $0.8\%$ when conditioned at $-5$ and $-10$\,dB, respectively. These results suggest that optimally selected or processed conditioning signals may further improve \ac{SE} performance. Overall, the model’s latent features appear to not directly represent clean speech or noise but rather more abstract non-stationary features that slightly emphasize speech characteristics.

\begin{figure}[!t]
  \centering
  % \vspace{-1.5em}
  % Adjust width as needed:
  \input{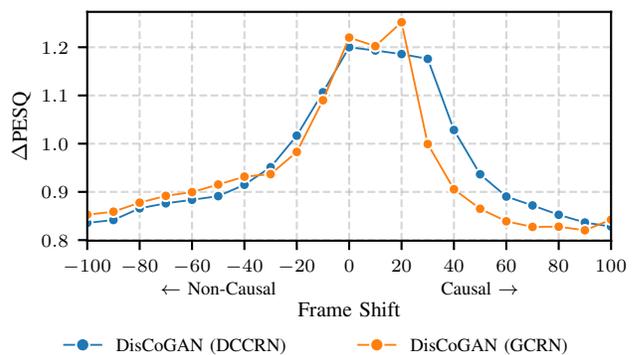}
  \vspace{-1em}
\caption{Effect of temporal misalignment of the latent features of the conditioning discriminative model on DisCoGAN performance.}
  \label{fig:frameshift}
  %  \vspace{-1em}
\end{figure}

\subsubsection{Robustness against Temporal Misalignment of Latent Features}
We evaluated the effect of shifting the latent features by \(k_1\geq 0\) frames on DisCoGAN on low SNR dataset. We applied causal (right) and non-causal (left) shifts to \(\Dl\):
\[
\begin{alignedat}{2}
\text{Causal:} \quad & \Dl[n, f] \leftarrow \Dl[n - k_1, f], \quad && \\
\text{Non-causal:} \quad & \Dl[n, f] \leftarrow \Dl[n + k_1, f].
\end{alignedat}
\]
Here, \(n\) and \(f\) are frame and feature indices. As shown in Fig.~\ref{fig:frameshift}, PESQ improvements for DisCoGAN (GCRN) and DisCoGAN (DCCRN) remain stable up to a 20–30 frame causal shift, suggesting DisCoGAN’s lookahead compensates for small delays. Non-causal shifts cause immediate degradation, e.g., a $10$-frame shift drops DisCoGAN (GCRN)  $\Delta$PESQ from $1.22$ to $1.08$. DisCoGAN (GCRN)  shows greater sensitivity to such shifts, likely due to differences in their STFT configurations (Tab.~\ref{Discriminative Models}); GCRN’s hop size aligns with the generator, whereas DCCRN operates at a higher resolution. Overall, neither model fails entirely under frame shifts, suggesting that the discriminative conditioning method uses both local and global temporal context rather than relying solely on frame-level alignment.

\section{Conclusions}
\label{conclusions}

In this work, we demonstrated the effectiveness of discriminative latent features as generic conditioning information to improve the performance of GAN-based \ac{SE} methods. Our proposed DisCoGAN method outperformed existing GAN-based techniques, including end-to-end, two-stage, and post-filtering methods—in low SNR scenarios, as well as state-of-the-art generative approaches in high SNR conditions and real-world scenarios. The ablation study showed that the proposed discriminative conditioning method
uses both local and global temporal context rather than
relying solely on frame-level alignment.

%Future work will focus on validating the effectiveness of this approach in related speech enhancement tasks such as joint acoustic echo cancellation and noise reduction, speech coding and enhancement, beamforming or embedding-based source extraction.

\section*{Acknowledgments}

The authors thank the Erlangen Regional Computing Center
(RRZE) for providing compute resources and technical support.

\bibliographystyle{IEEEbib}
\bibliography{strings}

\end{document}